\documentclass[%
 reprint,
 superscriptaddress,
 amsmath,amssymb,
 aps,
]{revtex4-1}

\usepackage{graphicx}
\usepackage{dcolumn}
\usepackage{bm}
\usepackage{color}
\usepackage{amssymb}
\usepackage{physics}
\usepackage{subfigure}

\begin{document}

\preprint{APS/123-QED}

\title{Efficient decoy states for the reference-frame-independent measurement-device-independent quantum key distribution}
	\author{Feng-Yu Lu}
	\author{Zhen-Qiang Yin}
	\email{yinzq@ustc.edu.cn}
    \author{Guan-Jie Fan-Yuan}
	\author{Rong Wang}
    \author{Hang Liu}
	\author{Shuang Wang}
	\author{Wei Chen}
	\author{De-Yong He}
	\affiliation{Key Laboratory of Quantum Information, CAS Center For Excellence in Quantum Information and Quantum Physics,
		University of Science and Technology of China, Hefei 230026, China}
	\affiliation{State Key Laboratory of Cryptology, P. O. Box 5159, Beijing 100878, P. R. China}
	\author{Wei Huang}
	\author{Bing-Jie Xu}
	\affiliation{Science and Technology on Communication Security Laboratory, Institute of Southwestern Communication, Chengdu, Sichuan 610041, China}
	\author{Guang-Can Guo}
	\author{Zheng-Fu Han}
	\affiliation{Key Laboratory of Quantum Information, CAS Center For Excellence in Quantum Information and Quantum Physics,
		University of Science and Technology of China, Hefei 230026, China}
	\affiliation{State Key Laboratory of Cryptology, P. O. Box 5159, Beijing 100878, P. R. China}




\date{\today}

\begin{abstract}
Reference-frame-independent measurement-device-independent quantum key distribution (RFI-MDI-QKD) is a novel protocol that eliminates all possible attacks on the detector side and necessity of reference-frame alignment in source sides. However, its performance may degrade notably due to statistical fluctuations, since more parameters, e.g. yields and error rates for mismatched-basis events, must be accumulated to monitor the security. In this work, we find that the original decoy states method estimates these yields over pessimistically since it ignores the potential relations between different bases. Through processing parameters of different bases jointly, the performance of RFI-MDI-QKD is greatly improved in terms of secret key rate and achievable distance when statistical fluctuations are considered (The statistical fluctuation analysis is the first step to the final analysis of finite key size effect). Our results pave an avenue towards practical RFI-MDI-QKD.
\end{abstract}

\pacs{Valid PACS appear here}
\maketitle

\section{introduction}
Quantum key distribution(QKD) \cite{BB84}, based on information-theoretic security guaranteed by quantum mechanics \cite{lo1999unconditional,shor2000simple,Scarani:QKDrev:2009,Rennersecurity}, allows two remote users, Alice and Bob, to share secret keys. In recent decades, numerous efforts have been made to improve the practical security and applicability of QKD protocols and systems \cite{hwang2003quantum,wang2005beating,lo2005decoy,boaron2018secure,yin2016measurement,wang20122,gordon2005quantum,wang2018practical,takesue2007quantum,tfqkd,lo2012measurement,acin2007device,sasaki2014practical,laing2010reference}. The measurement-device-independent quantum key distribution (MDI-QKD) \cite{lo2012measurement} and reference-frame-independent quantum key distribution (RFI-QKD) \cite{laing2010reference} are two of the most successful efforts of them. 

The MDI-QKD is proposed to eliminate all possible detector side channel attacks. In this protocol, the most fragile assumption \cite{fung2007phase,zhao2008quantum,gerhardt2011full,makarov2009controlling} that the measurement unit is ideal is removed. The protocol offers a great balance between practical security and usability. Up to now, several MDI-QKD experiment has been presented \cite{liu2013experimental,da2013proof,rubenok2013real,comandar2016quantum,yin2016measurement}. However, no matter which coding scheme, e.g. polarization coding or phase coding, was deployed, the coding reference frame of Alice and Bob have to be rigorous calibrated, which limits practical performance and may bring side channels to QKD systems.

On the other hand, the RFI-QKD allows Alice and Bob to share secret key bits without active alignment of reference-frame. It was pointed out that the RFI-QKD is particularly relevant for the earth to satellite QKD \cite{spedalieri2006quantum} and time-bin encoded QKD \cite{boaron2018secure}. However, the RFI-QKD still suffers from the fatal detector side-channel attacks \cite{fung2007phase,zhao2008quantum,gerhardt2011full,makarov2009controlling}.

Fortunately, the combination of MDI-QKD and RFI-QKD solve the above problems. The RFI-MDI-QKD protocol \cite{yin2014reference,wang2015phase} does not need active reference frame calibration and immune to all detector side channel attacks. The protocol simplifies the MDI-QKD, reduces the aligning expenses as well as security risk. Meanwhile, it closes all detector side loopholes of RFI-QKD, which makes the RFI-QKD more secure. 

Nevertheless, there are still two gaps between practice and theory. One of the gaps is the absence of ideal single-photon source. The gap is solved by decoy state method \cite{hwang2003quantum,wang2005beating,lo2005decoy,xu2014protocol,ma2012statistical,wang2013three,yu2015statistical,zhou2016making}. By randomly sending between so-called signal, decoy or vacuum states (weak coherent pulses with different intensities), Alice and Bob can establish a secret key strictly from information conveyed by single-photon component in weak coherent state. Another gap is named finite key size effect \cite{curty2014finite,sheridan2010finite}, which occurs when users only have limited resources while the theory assumes unlimited. The statistical fluctuation analysis \cite{ma2012statistical,yu2015statistical,zhou2016making} is the first step to the final analysis of finite key size effect. While the statistical fluctuation is considered, the secret key rate of RFI-MDI-QKD is rather undesirable. Compared with general MDI-QKD, the RFI version prepares more bases and more states, which means it requires more data to reduce the impact of the statistical fluctuations. For example, Refs. \cite{wang2017measurement,liu2018polarization} have experimentally demonstrated the RFI-MDI-QKD protocol to verify its feasibility, however, to reduce the impact of statistical fluctuations, the Ref. \cite{wang2017measurement} accumulates $3.5\times10^{11}$ pulse pairs for about 100 hours and the Ref. \cite{liu2018polarization} accumulates $3\times10^{12}$ pulse pairs for about 16 hours.
Since the RFI protocol assumes the misalignment of two unbiased bases are unknown but fixed, which may be spoiled in such a long system-running time and make the protocol not RFI any more. 

In this work, a new decoy state method for RFI-MDI-QKD is proposed. By considering the potential relations of different bases and applying the joint-study method \cite{yu2015statistical} our new method reaches much higher secret key rate and longer achievable distance in non-asymptotic scenarios compared with existing theoretical and experimental methods (our analysis and comparison only cotain statistical fluctuation but do not contain full finite key size effect analysis). Our new method can generate considerable key rate given a smaller number of trials, which means less time for data accumulation.

The remainder of this paper is organized as follows: In Sec. \ref{RFI_MDI_QKD}, we briefly introduce the RFI-MDI-QKD and the previous decoy state method \cite{wang2015phase,wang2017measurement,liu2018polarization}. Then in Sec. \ref{improved_protocol}, we present our improved protocol and introduce our theories and parameter estimation strategies for improving the performance in non-asymptotic cases. In Sec. \ref{section_simulation}, the simulation results are showed to verify the superiority of our new method. Finally, in Sec.\ref{discussion}, we summarize our new method and introduce a more intuitive explanation of the improvement.

\section{efficient decoy state RFI-MDI-QKD}

\subsection{original protocol}
\label{RFI_MDI_QKD}
We use subscript $A$ ($B$) to denote a variable or a parameter belongs to Alice (Bob). To simplicity, we will omit the subscript $A$ or $B$ if not causing any ambiguity.
We define $\mu$, $\nu$, $\omega$ and $o$ as intensities of weak coherent pulse in decoy state method. Especially, the $\nu>\omega>o$ and $o$ is the vacuum state. The probability of a weak coherent pulse with intensity $\lambda$ contains $k$ photons is $\lambda^k e^{-\lambda}/k!$, especially, we define $p_k^l = l^k e^{-l}/k!$ and ${p'}_k^r = r^k e^{-r}/k!$ are photon number distribution of Alice and Bob's weak coherent pulse respectively, where the $l$and $r$ denote the intensities selected by Alice and Bob respectively, the $k$ denotes the photon number.
The $\mathcal{X}$, $\mathcal{Y}$, $\mathcal{Z}$ are three coding bases. We define set $\mathcal{B} = \{ \mathcal{XX}, \mathcal{XY}, \mathcal{YX}, \mathcal{YY}, \mathcal{ZZ} \}$, the $b \in \mathcal{B}$ denotes the coding basis selected by Alice and Bob, especialy, the $\mathcal{S} = {\mathcal{ZZ}}$ is defined as both of the users select $\mathcal{Z}$ basis and $\mathcal{D} = \{ \mathcal{XX}, \mathcal{XY}, \mathcal{YX}, \mathcal{YY}\}$ is defined as a set which both of the users select $\mathcal{X}$ or $\mathcal{Y}$ basis. In original MDI-QKD, all bases should be carefully aligned, but in RFI-MDI-QKD, the misalignment of $\mathcal{X}$ and $\mathcal{Y}$ bases can be an unknown but fixed value, namely, the $\mathcal{X}$ and $\mathcal{Y}$ bases of Alice (Bob) can be changed as:
\begin{equation}
\begin{aligned}
&\mathcal{X}_{A(B)} = \mathcal{X}\cos{\beta_{A(B)}} + \mathcal{Y}\sin{\beta_{A(B)}};\\
&\mathcal{Y}_{A(B)} = \mathcal{Y}\cos{\beta_{A(B)}} - \mathcal{X}\sin{\beta_{A(B)}},
\label{reference_frame}
\end{aligned}
\end{equation}
where the $\beta$ denotes the reference frame misalignment.

The original decoy state RFI-MDI-QKD is described as follows:

\emph{Step 1. Preparation}: On each trial, Alice (Bob) randomly codes her (his) phase randomized weak coherent pulse on $\mathcal{Z}$ $\mathcal{X}$ or $\mathcal{Y}$ basis and randomly selects an intensity $l$ ($r$) from a pre-decided intensity set $\{ \nu_A, \omega_A, o \}$ ($\{ \nu_B, \omega_B, o \}$) \cite{wang2015phase,wang2017measurement,zhang2017practical,liu2018polarization}. After that, they send their weak coherent pulse to the untrusted third party Charlie. When both of Alice and Bob select intensity $\nu$ and $\mathcal{Z}$ basis, the pulse pair can be used to generate key bits.

\emph{Step 2. Measurement and announcement:} When charlie receives a weak coherent pulse pair 
from Alice and Bob, he does the Bell-state measurement and publicly announces his measurement result.

\emph{Step 3. Post-processing:} After the trial repeats $N_{tot}$-times, There are $N_{lr}^d$ pulse pairs with intensity pair $lr$ and basis pair $b$ in total. Alice and Bob can observe the overall gain and the overall error yield of each kind of pulse pair. We define $Q_{lr}^{d}$ and $T_{lr}^d$ as the overall gain and the overall error yield of the weak coherent pulse pairs with intensity pair $lr$ and basis pair $b$ respectively. Then, by applying the decoy state method \cite{wang2015phase,wang2017measurement,liu2018polarization,zhang2017practical,wang2013three}, they can estimate yield and quantum bit error rate of single-photon pulse pairs for each basis. We define $Y_{11}^d$ and $e_{11}^d$ as, respectively, the yield and quantum bit error rate of single-photon pulse pairs with basis pair $b$.
In asymptotic case, an observed value equals to its expected value. The lower bound of $Y_{11}^d$ and upper bound for each basis can be estimated by the decoy state method \cite{wang2013three,wang2015phase}.
\begin{equation}
\begin{aligned}
&Y_{11}^d \ge \underline{Y}_{11}^d = \big\{ \  [p_1^\nu {p'}_2^\nu\langle{Q}^d_{\omega\omega}\rangle + p_1^\omega {p'}_2^\omega p_0^\nu \langle{Q}^d_{o\nu}\rangle \\
& + p_1^\omega {p'}_2^\omega {p'}_0^\nu \langle{Q}^d_{\nu o}\rangle + p_1\nu {p'}_2^\nu p_0^\omega {p'}_0^\omega \langle{Q}_{oo}\rangle] - \ [ p_1^\omega {p'}_2^\omega \langle{Q}^d_{\nu\nu}\rangle \\
& + p_1^\nu {p'}_2^\nu p_0^\omega \langle{Q^d_{o\omega}}\rangle + p_1^\nu {p'}_2^\nu {p'}_0^\omega \langle{Q}^d_{\omega o}\rangle + p_0^\nu {p'}_0^\nu p_1^\omega {p'}_2^\omega \langle{Q}_{oo}\rangle\ ] \big\}\\
 & \big{/} [p_1^\nu p_2^\nu ( {p'}_1^\omega {p'}_2^\nu - {p'}_1^\nu {p'}_2^\omega )];\ \  \text{  for $d \in \mathcal{B}$},  
 \label{Y11_l}
\end{aligned}
\end{equation}
and $e_{11}^d \le \overline{e}_{11}^d = \overline{E}_{\lambda,11}^d / \underline{Y}_{11}^d$, where the $\lambda$ denotes the intensity of the decoy state. The $E_{\lambda,11}^d$ is defined as a intermediate variable, the upper bound $\overline{E}_{\lambda,11}^d$ can be estimated by:
\begin{equation}
\begin{aligned}
&\overline{E}_{\lambda,11}^{d} = \frac {[ \langle{T}_{\lambda}^{d}\rangle + p^{\lambda}_0{p'}^{\lambda}_0 \langle{T}_{oo}\rangle] - [p^{\lambda}_0 \langle{T}_{o\lambda}^d\rangle  + {p'}^{\lambda}_0 \langle{T}_{\lambda o}^d\rangle ]}{p^{\lambda}_1{p'}^{\lambda}_1},\\
& \text{  for $d \in \mathcal{B}$},  
\end{aligned}
\end{equation}

However, in non-asymptotic scenarios, the differences between observed values and expected values must be taken into considered. We define $\hat{Q}_{lr}^{d}$ and $\hat{T}_{lr}^{d}$ as the experimentally observed value, the $\langle{Q}_{lr}^{d}\rangle$ and $\langle{T}_{lr}^{d}\rangle$ as the expected value. In asymptotic case, the observed value equals to its expected value. In non-asymptotic cases, an expected value can be bounded by applying Chernoff bound on its observed value \cite{wang2017measurement,liu2018polarization,curty2014finite}: 
\begin{equation}
\begin{aligned}
&{N_{lr}^d}\langle {M_{lr}^d} \rangle \le
 \mathcal{F}^+(N_{lr}^d M_{lr}^d) =
 N_{lr}^d M_{lr}^d+
 f(\frac{(\frac{\epsilon}{2})^{4}}{16})\sqrt{N_{lr}^d M_{lr}^d } , \\
&{N_{lr}^d}\langle{M_{lr}^d}\rangle \ge 
\mathcal{F}^-(N_{lr}^d M_{lr}^d)=
N_{lr}^d M_{lr}^d -f((\frac{\epsilon}{2})^{\frac{3}{2}}) \sqrt{N_{lr}^d M_{lr}^d },
\end{aligned}
\label{Chernoff_bound}
\end{equation}

where $M_{lr}^{b}$ denotes $Q_{lr}^b$ or $T_{lr}^b$, the $\epsilon$ is the failure probability and $f(x) = \sqrt{2 \ln(x^{-1})}$. 

The estimation of $\underline{Y}_{11}^d$ and $\overline{E}_{\lambda,11}^d$ should be revised as follows:
\begin{equation}
\begin{aligned}
&\underline{Y}_{11}^d = \min: \\
&\big\{ \  [p_1^\nu {p'}_2^\nu {Q}^d_{\omega\omega} + p_1^\omega {p'}_2^\omega p_0^\nu {Q}^d_{o\nu} + p_1^\omega {p'}_2^\omega {p'}_0^\nu {Q}^d_{\nu o}\\
& + p_1^\nu {p'}_2^\nu p_0^\omega {p'}_0^\omega {Q}_{oo}]  -\ [p_1^\omega {p'}_2^\omega {Q}^d_{\nu\nu} + p_1^\nu {p'}_2^\nu p_0^\omega {Q^d_{o\omega}}\\
&  + p_1^\nu {p'}_2^\nu {p'}_0^\omega {Q}^d_{\omega o} + p_0^\nu {p'}_0^\nu p_1^\omega {p'}_2^\omega {Q}_{oo}] \big\} \big{/} [p_1^\nu p_2^\nu ( {p'}_1^\omega {p'}_2^\nu - {p'}_1^\nu {p'}_2^\omega )],  \\
&s.t. :\mathcal{F}^+(N_{lr}^b \hat{Q}_{lr}^b) \ge
  N_{lr}^b  Q_{lr}^b \ge
   \mathcal{F}^-(N_{lr}^b \hat{Q}_{lr}^b);\\
   & \ \text{for any $l,r \in \{\nu,\omega,o\}$},\\
 \label{Y11_l}
\end{aligned}
\end{equation}
and
\begin{equation}
\begin{aligned}
&\overline{E}_{\lambda,11}^{d} =max: \frac {[ {T}_{\lambda}^{d} + p^{\lambda}_0{p'}^{\lambda}_0 {T}_{oo}] - [p^{\lambda}_0 {T}_{o\lambda}^d  + {p'}^{\lambda}_0 {T}_{\lambda o}^d ]}{p^{\lambda}_1{p'}^{\lambda}_1},\\
&s.t. : \mathcal{F}^+{(N_{lr}^{d}\hat{T}_{lr}^{d})}\ge
 N_{lr}^{d}  T_{lr}^{d} \ge
 \mathcal{F}^-{(N_{lr}^{d}\hat{T}_{lr}^{d})};\\
 & \ \text{for any $l,r \in \{\lambda,o\}$}\\
\end{aligned}
\end{equation}
\cite{zhang2017practical,wang2017measurement,liu2018polarization}, where $d\in\mathcal{B}$ and $\lambda\in\{\nu,\omega\}$.

With above estimated yields and quantum bit error rates, the information leakage can be described as: 
\begin{equation}
I_{AE}(\underline{C},\overline{e}_{11}^\mathcal{S}) = (1 - \overline{e}_{11}^\mathcal{S})H_2[ (1+u)/2 ] + \overline{e}_{11}^\mathcal{S} H_2[(1+v)/2]
\label{I_AE}
\end{equation}
\cite{yin2014reference,laing2010reference}, where 
\begin{equation}
\underline{C} = \sum_{d\in\mathcal{D}} ( 1 - 2{\overline{e}_{11}^{d}})^2,
\label{quantity_c}
\end{equation}
and
\begin{equation}
\begin{aligned}
&u = \min{(\sqrt{\underline{C}/2}/(1-\overline{e}_{11}^\mathcal{S}),1)},\\
&v = \sqrt{\underline{C}/2 - (1 - \overline{e}_{11}^\mathcal{S})^2 u^2}/\overline{e}_{11}^\mathcal{S}.
\end{aligned}
\end{equation}
The $H_2(p) = -p\log_2p - (1 - p)log_2(1-p)$ is the Shannon entropy. The secret key rate can be estimated by the GLLP formula \cite{inamori2007unconditional,gottesman2004security}:
\begin{equation}
R \geq Pr_{\nu\nu}^\mathcal{S} p_1^{\nu}{p'}_1^{\nu} \big\{ \underline{Y}_{11}^\mathcal{S} \big[1 - I_{AE}(C,e_{11}^\mathcal{S})\big] - {Q^\mathcal{S}_{\nu\nu}}f_eH_2(\frac{T^\mathcal{S}_{\nu\nu}}{Q^\mathcal{S}_{\nu\nu}}) \big\},
\label{GLLP1}
\end{equation}
where the $Pr_{\nu\nu}^\mathcal{S}$ denotes the probability that Alice and Bob both select intensity $\nu$ and basis $Z$, the$Q^\mathcal{S}_{\nu\nu}$ and $T^\mathcal{S}_{\nu\nu}$ are overall gain and overall error rate when Alice and Bob select intensity pair $\nu\nu$ and basis pair $\mathcal{ZZ}$; the $f_e$ denotes the reconciliation efficiency in key reconciliation phase.


\begin{widetext}
\subsection{improved protocol}
\label{improved_protocol}

The statistical fluctuation seriously harms the secret key rate and achievable distance of QKD systems. Especially, compared with other types of QKD protocol, the statistical fluctuation harms the RFI-MDI-QKD system particularly serious since the RFI-MDI-QKD has more kinds of pulse pair which scatters the accumulated data and makes the fluctuation more obvious. Meanwhile, the $\beta$ may drift while accumulating data, which may invalidate the RFI protocol. 

In this work, we find that the previous works \cite{wang2015phase,wang2017measurement,liu2018polarization,zhang2017practical} estimate the yield and quantum bit error rate of each basis individually and ignore the potential relations of different basis. By analyzing these relations, we propose an improved decoy method for RFI-MDI-QKD which performs much better in non-asymptotic scenarios. Here we introduce our improved decoy state method based on potential relations between different bases for estimating reducing the impact of the statistical fluctuation.

In our new decoy state method, the \emph{step 1} is modified as follows:

\emph{New step 1 :} Alice (Bob) randomly selects $\mathcal{X}$, $\mathcal{Y}$, $\mathcal{Z}$ basis or selects vacuum state $o$ with no basis. If $\mathcal{Z}$ basis is selected, Alice (Bob) only select intensity $\mu_A$ ($\mu_B$); if $\mathcal{X}$ or $\mathcal{Y}$ basis is selected, Alice (Bob) randomly selects intensity $\nu_A$ or $\omega_A$ ($\nu_B$ or $\omega_B$). In the improved protocol, the weak coherent pulses in $\mathcal{Z}$ basis are only employed to generate key bits, we define the $\mathcal{Z}$ basis as the \emph{signal basis}; the $\mathcal{X}$ and $\mathcal{Y}$ bases are employed to estimated parameters, we define them as \emph{decoy basis}. Then, by applying three important relations between different bases, the performance of the protocol in non-asymptotic cases can be significantly improved.

\emph{Relation 1: Since the density matrices of single-photon pulse pair in $\mathcal{XX}$, $\mathcal{XY}$, $\mathcal{YX}$, $\mathcal{YY}$ and $\mathcal{ZZ}$ bases are quite the same, the yield $Y_{11}^\mathcal{XX}=Y_{11}^\mathcal{XY}=Y_{11}^\mathcal{YX}=Y_{11}^\mathcal{YY}=Y_{11}^{\mathcal{S}}$ }\cite{zhou2016making} (We emphasize that these $Y_{11}^d$ are ideal yield of single-photon pulse pairs, namely, the yield when Alice and Bob have single-photon source and the number of trials is infinite. They are independent of decoy state method and statistical fluctuation).

Based on the \emph{Relation 1}, we can estimate the $Y_{11}^\mathcal{S}$ by $\mathcal{X}$ or $\mathcal{Y}$ basis, namely, the decoy basis. In signal basis $\mathcal{Z}$, we only modulate intensity $\mu$ to generate key bits. 

\emph{Relation 2: When estimating $Y_{11}^\mathcal{S}$, all pulse pairs in decoy basis can be regarded as an entirety. The set $\mathcal{D}$ include all decoy basis, define $Y_{11}^\mathcal{D}$ as the average yield of single-photon pulse pairs in the joint decoy basis. It is obvious that, the $Y_{11}^\mathcal{S}$ = $Y_{11}^\mathcal{D}$. To simplification, we use $Y_11$ to denote the yield of single-photon pulse pairs for any basis.}


We define $ Q_{lr}^\mathcal{D}  = \frac{\sum_{d\in\mathcal{D}}N_{lr}^{d}  Q_{lr}^{d} }{\sum_{d\in\mathcal{D}}N_{lr}^d}$ as the gain of the weak coherent pulse pair in the joint decoy basis. By applying the \emph{relation 1}, \emph{relation 2} and the joint-study method proposed in Ref. \cite{yu2015statistical}, we can estimate a tight common lower bound of $Y_{11}$ for any of the basis
by solving the linear programming

\begin{equation}
\begin{aligned}
&  \underline{Y}_{11} =
\min:\ \big\{ \  [p_1^\nu {p'}_2^\nu {Q}^\mathcal{D}_{\omega\omega} + p_1^\omega {p'}_2^\omega p_0^\nu {Q}^\mathcal{D}_{o\nu} + p_1^\omega {p'}_2^\omega {p'}_0^\nu {Q}^\mathcal{D}_{\nu o} + p_1^\nu {p'}_2^\nu p_0^\omega {p'}_0^\omega {Q}_{oo}] \\- 
 & \ [p_1^\omega {p'}_2^\omega {Q}^\mathcal{D}_{\nu\nu} + p_1^\nu {p'}_2^\nu c_0^\omega {Q^\mathcal{D}_{o\omega}} + p_1^\nu {p'}_2^\nu {p'}_0^\omega {Q}^\mathcal{D}_{\omega o} + p_0^\nu {p'}_0^\nu p_1^\omega {p'}_2^\omega {Q}_{oo}] \big\} \big{/} [p_1^\nu p_2^\nu( {p'}_1^\omega {p'}_2^\nu - {p'}_1^\nu {p'}_2^\omega )], \\
 s.t.:&\\
 &\mathcal{F}^+(N_{lr}^\mathcal{D}\hat{Q}_{lr}^\mathcal{D}) \ge
  N_{lr}^\mathcal{D} Q_{lr}^\mathcal{D}  \ge
   \mathcal{F}^-(N_{lr}^\mathcal{D}\hat{Q}_{lr}^\mathcal{D}); \ \text{for any $l,r \in \{\nu,\omega,o\}$}, \\
 &\mathcal{F}^+(N_{\nu o}^\mathcal{D} \hat{Q}_{\nu o}^\mathcal{D}  + N_{\nu o}^\mathcal{D} \hat{Q}_{\nu o}^\mathcal{D} ) \ge
  N_{\nu o}^\mathcal{D} Q_{\nu o}^\mathcal{D}  + N_{\nu o}^\mathcal{D} Q_{\nu o}^\mathcal{D}  \ge
   \mathcal{F}^-(N_{\nu o}^\mathcal{D} \hat{Q}_{\nu o}^\mathcal{D} + N_{\nu o}^\mathcal{D} \hat{Q}_{\nu o}^\mathcal{D} ), \\
 &\mathcal{F}^+(N_{\omega o}^\mathcal{D} \hat{Q}_{\omega o}^\mathcal{D}  + N_{\omega o}^\mathcal{D} \hat{Q}_{\omega o}^\mathcal{D} ) \ge
  N_{\omega o}^\mathcal{D} Q_{\omega o}^\mathcal{D}  + N_{\omega o}^\mathcal{D} Q_{\omega o}^\mathcal{D} \ge
   \mathcal{F}^-(N_{\omega o}^\mathcal{D} \hat{Q}_{\omega o}^\mathcal{D}  + N_{\omega o}^\mathcal{D} \hat{Q}_{\omega o}^\mathcal{D} ), \\
 &  
  N_{\nu \nu}^\mathcal{D} Q_{\nu \nu}^\mathcal{D}  +
  N_{\nu o}^\mathcal{D} Q_{\nu o}^\mathcal{D}  + 
  N_{o\nu}^\mathcal{D} Q_{o\nu}^\mathcal{D}  +
  N_{o o}  Q_{o o}  \le 
  \mathcal{F}^+( N_{\nu \nu}^\mathcal{D} \hat{Q}_{\nu \nu}^\mathcal{D}  +
  N_{\nu o}^\mathcal{D} \hat{Q}_{\nu o}^\mathcal{D} + 
  N_{o\nu}^\mathcal{D} \hat{Q}_{o\nu}^\mathcal{D}  +
  N_{o o} \hat{Q}_{o o})\\
 &  N_{\nu \nu}^\mathcal{D} Q_{\nu \nu}^\mathcal{D}  +
  N_{\nu o}^\mathcal{D} Q_{\nu o}^\mathcal{D}  + 
  N_{o\nu}^\mathcal{D} Q_{o\nu}^\mathcal{D}  +
  N_{o o} Q_{o o}  \ge
  \mathcal{F}^-( N_{\nu \nu}^\mathcal{D} \hat{Q}_{\nu \nu}^\mathcal{D}  +
  N_{\nu o}^\mathcal{D} \hat{Q}_{\nu o}^\mathcal{D}  + 
  N_{o\nu}^\mathcal{D} \hat{Q}_{o\nu}^\mathcal{D}  +
  N_{o o} \hat{Q}_{o o} ), \\
 & N_{\omega \omega}^\mathcal{D} Q_{\omega \omega}^\mathcal{D}  +
  N_{\omega o}^\mathcal{D} Q_{\omega o}^\mathcal{D}  + 
  N_{o\omega}^\mathcal{D} Q_{o\omega}^\mathcal{D}  +
  N_{o o} Q_{o o}  \le
  \mathcal{F}^+( N_{\omega \omega}^\mathcal{D} \hat{Q}_{\omega \omega}^\mathcal{D}  +
  N_{\omega o}^\mathcal{D} \hat{Q}_{\omega o}^\mathcal{D}  + 
  N_{o\omega}^\mathcal{D} \hat{Q}_{o\omega}^\mathcal{D}  +
  N_{o o} \hat{Q}_{o o} ),\\
 &  
  N_{\omega \omega}^\mathcal{D} Q_{\omega \omega}^\mathcal{D}  +
  N_{\omega o}^\mathcal{D} Q_{\omega o}^\mathcal{D}  + 
  N_{o\omega}^\mathcal{D} Q_{o\omega}^\mathcal{D}  +
  N_{o o} Q_{o o}  \ge
  \mathcal{F}^-( N_{\omega \omega}^\mathcal{D} \hat{Q}_{\omega \omega}^\mathcal{D}  +
  N_{\omega o}^\mathcal{D} \hat{Q}_{\omega o}^\mathcal{D}  + 
  N_{o\omega}^\mathcal{D} \hat{Q}_{o\omega}^\mathcal{D}  +
  N_{o o} \hat{Q}_{o o} ).
 \label{Y11_l}
\end{aligned}
\end{equation}
The first constraint is obtained by applying Chernoff bound on experimental observed values and the other constraints are obtained by the joint-study method proposed by Ref. \cite{yu2015statistical}.
\end{widetext} 	


\begin{widetext}
\emph{Relation 3: Any combinations of $\mathcal{XX}$, $\mathcal{XY}$, $\mathcal{YX}$ and $\mathcal{YY}$ basis can be regarded as an entirety. Define $e_{11}^{d_1d_2}$ is the quantum bit error rate of the single-photon pulse pairs in the joint basis of $d_1$ and $d_2$, where $d_1,d_2 \in \mathcal{D}$ and $d_1 \neq d_2$. Define $e_{11}^{\mathcal{D}}$ is the quantum bit error rate of the single-photon pulse pairs the joint basis $\mathcal{D}$. When Alice and Bob select $\mathcal{X}$ and $\mathcal{Y}$ basis with equal probability, the $e_{11}^{d_1d_2} = ({e_{11}^{d_1}+e_{11}^{d_2}})/{2}$, the $e_{11}^{\mathcal{D}} = ({\sum_{d\in\mathcal{D}}{e_{11}^{d}} })/{4}$ 
}

Similar to the definition of $Q_{lr}^\mathcal{D}$, we define $T_{lr}^{{d_1}{d_2}} = \frac{T_{lr}^{d_1} N_{lr}^{d_1}+T_{lr}^{d_2} N_{lr}^{d_2}}{N_{lr}^{d_1}+N_{lr}^{d_2}}$ for $d_1,d_2 \in\mathcal{D}$ and $d_1 \neq d_2 $ and $T_{lr}^\mathcal{D}  = \frac{\sum_{d\in\mathcal{D}}N_{lr}^{d}  T_{lr}^{d} }{\sum_{d\in\mathcal{D}}N_{lr}^d}$.
By applying the \emph{Relation 3}, we can introduce more constraints for estimating a tighter lower bound of information leakage. Similar to the definition of the ${E}_{\lambda,11}^{d}$, we define the intermediate variables ${E}_{\lambda,11}^{{d_1}{d_2}} = Y_{11}e_{11}^{{d_1}{d_2}}$ and ${E}_{\lambda,11}^\mathcal{D}=Y_{11}e_{11}^\mathcal{D}$, the upper bound of these ${E}_{\lambda,11}$ can be estimated by linear programming and the joint-study method too:
\begin{equation}
\begin{aligned}
&\overline{E}_{\lambda,11}^{d} = \max : \frac {[ T_{\lambda\lambda}^{d} + p_0^\lambda {p'}^\lambda_0 T_{oo}] - [p^\lambda_0 T_{o\lambda}^{d}  + {p'}^\lambda_0 T_{\lambda o}^{d} ]}{p^\lambda_1{p'}^\lambda_1};\  \text{for $\lambda \in \{\nu,\omega \} $ if $d \in \mathcal{D}$ and $\lambda = \{\mu\} $ if $d = \mathcal{S}$  }, \\
&s.t.:\\
&\mathcal{F}^+{(N_{lr}^{d}
\hat{T}_{lr}^{d})}\ge
 N_{lr}^{d}  T_{lr}^{d}   \ge
 \mathcal{F}^-{(N_{lr}^{d}\hat{T}_{lr}^{d})};  \ \text{for any $l,r \in \{\lambda,o\}$};  \ \text{for any $l,r \in \{\lambda,o\}$}\\
 &\mathcal{F}^+{(N_{\lambda o}^{d}\hat{T}_{\lambda o}^{d}+N_{ o\lambda}^{d}\hat{T}_{o\lambda}^{d} )  }\ge
 N_{\lambda o}^{d}  T_{\lambda o}^{d}  + N_{ o\lambda}^{d}  T_{ o\lambda}^{d}   \ge
 \mathcal{F}^-{(N_{\lambda o}^{d}\hat{T}_{\lambda o}^{d} + N_{o\lambda}^{d}\hat{T}_{o\lambda}^{d})}\  ,\\
 &\mathcal{F}^+{(N_{\lambda \lambda}^{d}\hat{T}_{\lambda \lambda}^{d}+N_{ oo}\hat{T}_{oo} )  }\ge
 N_{\lambda \lambda}^{d}  T_{\lambda \lambda}^{d}  + N_{oo}  T_{oo}   \ge
 \mathcal{F}^-{(N_{\lambda \lambda}^{d}\hat{T}_{\lambda \lambda}^{d} + N_{oo}\hat{T}_{oo})}\ .\\
\end{aligned}
\end{equation}

\begin{equation}
\begin{aligned}
&\overline{E}_{\lambda,11}^{d_1 d_2} = \max : \frac {[ T_{\lambda\lambda}^{d_1d_2} + p_0^\lambda {p'}^\lambda_0 T_{oo}] - [p^\lambda_0 T_{o\lambda}^{ d_1 d_2}  + {p'}^\lambda_0 T_{\lambda o}^{d_1 d_2} ]}{p^\lambda_1{p'}^\lambda_1};\  \text{for $\lambda \in \{\nu,\omega \} $, $d_1,d_2 \in \mathcal{D}$ and $d_1\neq d_2$}, \\
&s.t.:\\
&\mathcal{F}^+{(N_{lr}^{d_1 d_2}\hat{T}_{lr}^{d_1 d_2})}\ge
 N_{lr}^{d_1 d_2}  T_{lr}^{d_1 d_2}   \ge
 \mathcal{F}^-{(N_{lr}^{d_1 d_2}\hat{T}_{lr}^{d_1 d_2})};  \ \text{for any $l,r \in \{\lambda,o\}$},\\
 &\mathcal{F}^+{(N_{\lambda o}^{d}\hat{T}_{\lambda o}^{d_1d_2}+N_{ o\lambda}^{d_1d_2}\hat{T}_{o\lambda}^{d_1d_2} )  }\ge
 N_{\lambda o}^{d_1d_2}  T_{\lambda o}^{d_1d_2}  + N_{ o\lambda}^{d_1d_2}  T_{ o\lambda}^{d_1d_2}   \ge
 \mathcal{F}^-{(N_{\lambda o}^{d_1d_2}\hat{T}_{\lambda o}^{d_1d_2} + N_{o\lambda}^{d_1d_2}\hat{T}_{o\lambda}^{d_1d_2})}\  ,\\
 &\mathcal{F}^+{(N_{\lambda \lambda}^{d_1d_2}\hat{T}_{\lambda \lambda}^{d_1d_2}+N_{ oo}\hat{T}_{oo} )  }\ge
 N_{\lambda \lambda}^{\mathcal{D}}  T_{\lambda \lambda}^{d_1d_2}  + N_{ oo}  T_{ oo}   \ge
 \mathcal{F}^-{(N_{\lambda \lambda}^{d_1d_2}\hat{T}_{\lambda \lambda}^{d_1d_2} + N_{oo}\hat{T}_{oo})}\ .\\
\end{aligned}
\end{equation}

\begin{equation}
\begin{aligned}
&\max : \overline{E}_{\lambda,11}^{\mathcal{D}} = \frac {[ T_{\lambda\lambda}^{\mathcal{D}} + p_0^\lambda {p'}^\lambda_0 T_{oo}] - [p^\lambda_0 T_{o\lambda}^{\mathcal{D}}  + {p'}^\lambda_0 T_{\lambda o}^{\mathcal{D}} ]}{p^\lambda_1{p'}^\lambda_1};\  \text{for $\lambda \in \{\nu,\omega \} $}, \\
&s.t.:\\
&\mathcal{F}^+{(N_{lr}^{\mathcal{D}}\hat{T}_{lr}^{\mathcal{D}})}\ge
 N_{lr}^{\mathcal{D}}  T_{lr}^{\mathcal{D}}   \ge
 \mathcal{F}^-{(N_{lr}^{\mathcal{D}}\hat{T}_{lr}^{\mathcal{D}})};  \ \text{for any $l,r \in \{\lambda,o\}$},\\
 &\mathcal{F}^+{(N_{\lambda o}^{\mathcal{D}}\hat{T}_{\lambda o}^{\mathcal{D}}+N_{ o\lambda}^{\mathcal{D}}\hat{T}_{o\lambda}^{\mathcal{D}} )  }\ge
 N_{\lambda o}^{\mathcal{D}}  T_{\lambda o}^{\mathcal{D}}  + N_{ o\lambda}^{\mathcal{D}}  T_{ o\lambda}^{\mathcal{D}}   \ge
 \mathcal{F}^-{(N_{\lambda o}^{\mathcal{D}}\hat{T}_{\lambda o}^{\mathcal{D}} + N_{o\lambda}^{\mathcal{D}}\hat{T}_{o\lambda}^{\mathcal{D}})}\  ,\\
 &\mathcal{F}^+{(N_{\lambda \lambda}^{\mathcal{D}}\hat{T}_{\lambda \lambda}^{\mathcal{D}}+N_{ oo}\hat{T}_{oo} )  }\ge
 N_{\lambda \lambda}^{\mathcal{D}}  T_{\lambda \lambda}^{\mathcal{D}}  + N_{ oo}^{\mathcal{D}}  T_{ oo}   \ge
 \mathcal{F}^-{(N_{\lambda \lambda}^{\mathcal{D}}\hat{T}_{\lambda \lambda}^{\mathcal{D}} + N_{oo}\hat{T}_{oo})}\ .\\
\end{aligned}
\end{equation}

With above definitions, a tighter lower bound for the intermediate variable $\underline{C}$ can be estimated by optimization algorithms:
\begin{equation}
\begin{aligned}
& \underline{C} = \min :\sum_{d\in \mathcal{D}} {(1 - 2e^d_{11})^2},\\
&s.t.:\\
&e_{11}^{{d}} \ge 0,\\
&e_{11}^{{d}} \le  \min({\overline{E}_{\nu,11}^{d}},{\overline{E}_{\omega,11}^{d}})/{\underline{Y}_{11}},\\
&e_{11}^{{d}_1} + e_{11}^{{d}_2} \le 2\min({\overline{E}_{\nu,11}^{d_1d_2}},{\overline{E}_{\omega,11}^{d_1d_2}})/\underline{Y}_{11},\  \\
&\sum_{d\in\mathcal{D}}{e_{11}^d} \le 4\min({\overline{E}_{\nu,11}^{\mathcal{D}}},{\overline{E}_{\omega,11}^{\mathcal{D}}})/\underline{Y}_{11};\\
&\text{for $d,d_1,d_2 \in \mathcal{D} $ and $d_1 \neq d_2$}.
\end{aligned}
\end{equation}

Finally, a tighter secret key rate in non-asymptotic scenario can be estimated as follows:
\begin{equation}
R \geq Pr_{\mu\mu}^\mathcal{S} p_1^\mu {p'}_1^\mu \big\{\underline{Y}_{11} \big[1 - I_{AE}(\underline{C},\overline{e}_{11}^\mathcal{S})\big] - {Q^\mathcal{S}_{\mu\mu}}fH_2(\frac{T^\mathcal{S}_{\mu\mu}}{Q^\mathcal{S}_{\mu\mu}})\big\},
\label{GLLP2}
\end{equation}
where
$Pr_{\mu\mu}^\mathcal{S}$ denotes the probability that Alice and Bob both select the code mode, the $\overline{e}_{11}^\mathcal{S} = \overline{E}_{\mu,11}^\mathcal{S}/\underline{Y}_{11}$ and the intermediate values are:
\begin{equation}
\begin{aligned}
\\
&v = \sqrt{\underline{C}/2 - (1 - \overline{e}_{11}^\mathcal{S})^2 u^2}/\overline{e}_{11}^\mathcal{S},\\
&u = \min{(\sqrt{\underline{C}/2}/(1-\overline{e}_{11}^\mathcal{S}),1)},\\
&I_{AE}(\underline{C},\overline{e}_{11}^\mathcal{S}) = (1 - \overline{e}_{11}^\mathcal{S})H_2[ (1+u)/2 ] + \overline{e}_{11}^\mathcal{S} H_2[(1+v)/2].
\label{I_AE}
\end{aligned}
\end{equation}

In the end of this section, we briefly summarize the difference between our improved method and the previous method. Our new method introduces the four intensity $\mu$ which is employed to generate key bits and the intensities $\nu$, $\omega$ and $o$ are only employed to estimate parameters. The new method also introduce three potential relations of each modes and each basis. By applying these ralations and the joint-study method, the impact of the statistical fluctuation is reduced. Thus our method can significantly improve the performance of the protocol in non-asymptotic cases.
\end{widetext}


\section{simulation}
\label{section_simulation}
In this section, we simulate our new decoy state method for RFI-MDI-QKD and compare the results with the best known prior article results (Refs.\cite{zhang2017practical,liu2018polarization}) with device parameters listed in Tab.\ref{device_parameters}. (The simulations only contain the statistical fluctuation analysis but do not contain the complete analysis of finite key size effect.)

\begin{figure}[htbp]
\includegraphics[width=9cm]{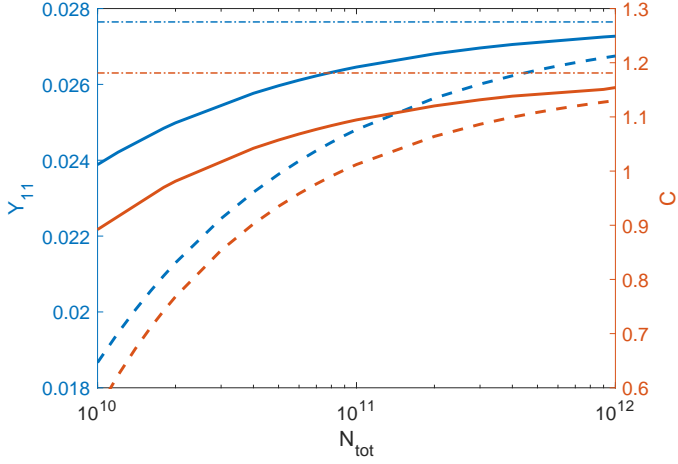}
\caption{\label{estimate_YandC} The estimated $\underline{Y}_{11}$ and $\underline{C}$ as a function of data size. The blue and red line denote the estimated value of $\underline{Y}_{11}$ and $\underline{C}$ respectively; the solid line denotes our improved method and the dash line denotes the original method; the dash-dot line denotes the asymptotic cases. The $\nu$ and $\omega$ are fixed to 0.2 and 0.05 respectively, the probability of selecting $\mu$, $\nu$, $\omega$ and $o$ are equal; the device parameters are listed in Tab. \ref{device_parameters}}
\end{figure}

Firstly, we illustrate the improvement of estimating the $\underline{Y}_{11}$ and $\underline{C}$ in Fig. \ref{estimate_YandC}. By employing our \emph{relation 2} and \emph{relation 3}, the $\underline{Y}_{11}$ and $\underline{C}$ are significantly estimated tighter, especially when $N_{tot}$ is less than $10^{11}$. 

\begin{table}[b]
\caption{\label{device_parameters}
The table lists the device parameters of our simulation. The $\eta_d$ denotes the detection efficiency; the ${p_d}$ denotes the dark count rate; the $e_d$ denotes the misalignment error rate; the $f_e$ denotes the reconciliation efficiency in key reconciliation phase; the $\epsilon$ denotes the failure probability of parameter estimation; the $\alpha$ denotes the fiber loss per km.}
\begin{ruledtabular}
\begin{tabular}{cccccc}
   $\eta_d$  & ${p_d}$ &  $e_d$ &  $f_e$ & $\epsilon$ & $\alpha$ \\
\hline
  $25\%$ & $10^{-6}$ & $0.5\%$ & $1.16$ & $10^{-7}$ & $0.2$ dB/km 
\end{tabular}
\end{ruledtabular}
\end{table}

Then we simulate the optimized secret key rate as a function of transmission distance.
We focus on the symmetric case where the distance from Alice to Charlie and from Bob to Charlie are equal, due to the symmetric case, we treat the parameters of Alice Bob equivalently for simplicity. We define $\mu = \mu_A =\mu_B$, $\nu = \nu_A =\nu_B$, $\omega = \omega_A =\omega_B$ and define $Pr^\mathcal{S}_\mu$, $Pr^\mathcal{D}_\nu$, $Pr^\mathcal{D}_\omega$ as the probability of Alice (Bob) selects intensity $\mu$ $\nu$ $\omega$ respectively. The probability of selecting vacuum state $o$ is $1 - Pr^\mathcal{S}_\mu - Pr^\mathcal{D}_\nu - Pr^\mathcal{D}_\omega$. Alice (Bob) selects $\mathcal{X}$ or $\mathcal{Y}$ basis with equal probability. By optimizing $\mu$, $\nu$, $\omega$, $Pr^\mathcal{S}_\mu$, $Pr^\mathcal{D}_\nu$, $Pr^\mathcal{D}_\omega$ with particle swarm optimization algorithm (PSO) \cite{shi1999empirical}, we investigate our new decoy method at different data size $N_{tot}$ and compare our result with previous works \cite{zhang2017practical,liu2018polarization}. 

\begin{figure}[htbp]
\includegraphics[width=9cm]{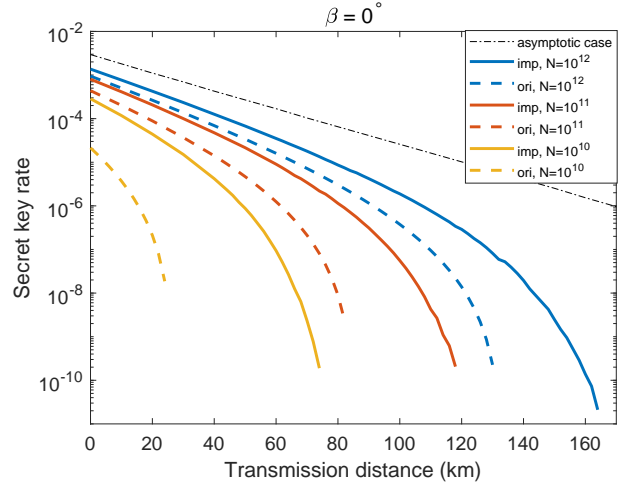}
\caption{\label{R_L_beta0} The secret key rate as function of communication distance. In this figure, the reference frame of Alice and Bob is well aligned.
 The black dash-dot line denotes the asymptotic case. The solid line and dash line denote, respectively, our improved decoy method and original method; the blue, red and yellow denote, respectively, $N_{tot}$ is $10^{12}$, $10^{11}$ and $10^{10}$.}
\end{figure}

\begin{figure}[htbp]
\includegraphics[width=9cm]{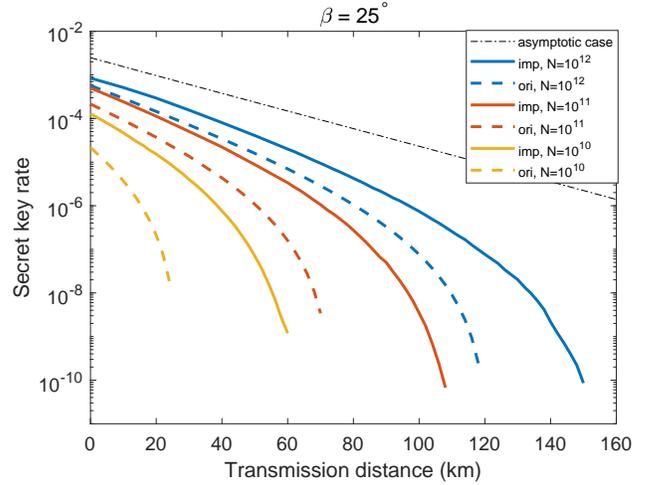}
\caption{\label{R_L_beta25} The secret key rate as function of communication distance. In this figure, the reference frame misalignment of Alice and Bob is fixed to $25^\circ$.
 The black dash-dot line denotes the asymptotic case. The solid line and dash line denote, respectively, our improved decoy method and original method; the blue, red and yellow denote, respectively, $N_{tot}$ is $10^{12}$, $10^{11}$ and $10^{10}$.}
\end{figure}

In Fig. \ref{R_L_beta0} and Fig. \ref{R_L_beta25}, the comparison results at different reference frame misalignment are illustrated. The simulation results indicate that our new method and the previous method have the same performance in the asymptotic case, but when data size is finite, our method performs much better.
Our method can generate considerable key rate with about only $10^{10}$ data size, which is significantly less than previous works. When $50$ MHz system \cite{liu2018polarization} is employed, only several minutes are needed to accumulate enough data. Especially, if the GHz high-speed system \cite{thew2006low,takesue2007quantum,wang20122} system is employed, only several seconds are needed, which can make the protocol practically reference frame independent. 

\begin{figure}[htbp]
\includegraphics[width=9cm]{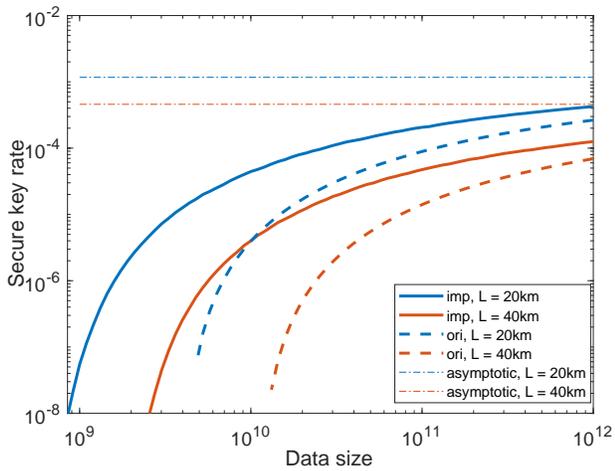}
\caption{\label{R_N_beta0.eps} The secert key rate versus different data size. The reference-frame misalignment $\beta$ is fixed to $0$ here. The blue and red line denote, respectively, transmission distance is 20 and 40 km. The solid line denotes our new method, the dash line denotes the previous method and the dash-dot lines are the asymptotic cases. Especially, when $N_{tot}>10^{17}$, the secret key rate nearly reaches the asymptotic case. }
\end{figure}

The Fig. \ref{R_N_beta0.eps} illustrate the secret key rate as a function of data size $N_{tot}$. The simulation results indicate that our improved method can generate considerable key rate when $N_{tot} = 10^{10}$ while the original method need more than $10^{11}$ or $10^{12}$. When the GHz system is employed, the $N_{tot} = 10^{10}$ is an accepted data size to ensure the system practical RFI. In the scenario of $N_{tot} = 10^{10}$, $\beta=0$ and $L = 10$-km, the secret key rate of our method is $130$ times of previous method; when $L = 20$ km, our method is $3000$ times of previous.

\section{discussion}
\label{discussion}
In the practical RFI-MDI-QKD systems, the statistical fluctuation must be taken into considered. 
The previous works \cite{wang2017measurement,zhang2017practical,liu2018polarization} ignore the potential relations of each modes and each basis and only use the worst-case calculation for estimating $\underline{Y}_{11}$ and $\underline{C}$.
To alleviate the impact of statistical fluctuation, a new decoy state method for RFI-MDI-QKD is proposed. Our method employs three relations introduced in Sec. \ref{improved_protocol} and the joint-study method proposed in Ref. \cite{yu2015statistical}; we only modulate intensity $\mu$ in $\mathcal{Z}$ basis and estimates the yield by the combination of decoy basis. We also introduce more constraints for the intermediate variable $C$, which makes the information leakage estimated much tighter. 
The simulation results indicate that our method can generate a considerable secret key rate with only $10^{10}$ data size, which is significantly less than the previous method. Our method significantly reduces the time for accumulating data, which mitigates the drift of the reference frame during accumulating data.

Here, we introduce a more intuitive explanation for our improvement. Firstly, in previous works, the signal state is used to generate key rates and parameter estimation. However, in non-asymptotic cases, the optimized intensity for generating key rates is usually not equal to it for parameter estimation. In our new method, we introduce the fourth intensity for parameter estimation and the signal state is not used to estimate $Y_{11}$ anymore, which leads to a higher secret key rate when all parameters are optimized. Secondly, the new method only modulates $\mu$ on $\mathcal{Z}$ basis, which means less observed values are needed. Finally, by considering the potential relations of different bases and different modes, more constraints can be introduced when estimating $\underline{Y}_{11}$ and $\underline{C}$, which makes them estimated much tighter. By applying our improved decoy stated method, the performance of RFI-MDI-QKD can be greatly improved and the time for data accumulation can be significantly reduced. Our results pave an avenue towards practical RFI-MDI-QKD. 

\section{ACKNOWLEDGMENTS}
We gratefully appreciate the help from Xing-Yu Zhou and Chun-Mei Zhang.

This work has been supported by the National Key Research and Development Program of China (Grant No. 2016YFA0302600), the National Natural Science Foundation of China (Grant Nos. 61822115, 61961136004, 61775207, 61702469, 61771439, 61627820), National Cryptography Development Fund (Grant No. MMJJ20170120) and Anhui Initiative in Quantum Information Technologies.

\nocite{*}

\bibliography{apssamp}

\begin{thebibliography}{49}%
\makeatletter
\providecommand \@ifxundefined [1]{%
 \@ifx{#1\undefined}
}%
\providecommand \@ifnum [1]{%
 \ifnum #1\expandafter \@firstoftwo
 \else \expandafter \@secondoftwo
 \fi
}%
\providecommand \@ifx [1]{%
 \ifx #1\expandafter \@firstoftwo
 \else \expandafter \@secondoftwo
 \fi
}%
\providecommand \natexlab [1]{#1}%
\providecommand \enquote  [1]{``#1''}%
\providecommand \bibnamefont  [1]{#1}%
\providecommand \bibfnamefont [1]{#1}%
\providecommand \citenamefont [1]{#1}%
\providecommand \href@noop [0]{\@secondoftwo}%
\providecommand \href [0]{\begingroup \@sanitize@url \@href}%
\providecommand \@href[1]{\@@startlink{#1}\@@href}%
\providecommand \@@href[1]{\endgroup#1\@@endlink}%
\providecommand \@sanitize@url [0]{\catcode `\\12\catcode `\$12\catcode
  `\&12\catcode `\#12\catcode `\^12\catcode `\_12\catcode `\%12\relax}%
\providecommand \@@startlink[1]{}%
\providecommand \@@endlink[0]{}%
\providecommand \url  [0]{\begingroup\@sanitize@url \@url }%
\providecommand \@url [1]{\endgroup\@href {#1}{\urlprefix }}%
\providecommand \urlprefix  [0]{URL }%
\providecommand \Eprint [0]{\href }%
\providecommand \doibase [0]{http://dx.doi.org/}%
\providecommand \selectlanguage [0]{\@gobble}%
\providecommand \bibinfo  [0]{\@secondoftwo}%
\providecommand \bibfield  [0]{\@secondoftwo}%
\providecommand \translation [1]{[#1]}%
\providecommand \BibitemOpen [0]{}%
\providecommand \bibitemStop [0]{}%
\providecommand \bibitemNoStop [0]{.\EOS\space}%
\providecommand \EOS [0]{\spacefactor3000\relax}%
\providecommand \BibitemShut  [1]{\csname bibitem#1\endcsname}%
\let\auto@bib@innerbib\@empty
\bibitem [{\citenamefont {Bennett}\ and\ \citenamefont
  {Brassard}(1984)}]{BB84}%
  \BibitemOpen
  \bibfield  {author} {\bibinfo {author} {\bibfnamefont {C.~H.}\ \bibnamefont
  {Bennett}}\ and\ \bibinfo {author} {\bibfnamefont {G.}~\bibnamefont
  {Brassard}},\ }in\ \href@noop {} {\emph {\bibinfo {booktitle} {Proceedings of
  IEEE International Conference on Computers, Systems and Signal Processing}}}\
  (\bibinfo  {publisher} {IEEE},\ \bibinfo {year} {1984})\ pp.\ \bibinfo
  {pages} {175--179}\BibitemShut {NoStop}%
\bibitem [{\citenamefont {Lo}\ and\ \citenamefont
  {Chau}(1999)}]{lo1999unconditional}%
  \BibitemOpen
  \bibfield  {author} {\bibinfo {author} {\bibfnamefont {H.-K.}\ \bibnamefont
  {Lo}}\ and\ \bibinfo {author} {\bibfnamefont {H.~F.}\ \bibnamefont {Chau}},\
  }\href@noop {} {\bibfield  {journal} {\bibinfo  {journal} {science}\ }\textbf
  {\bibinfo {volume} {283}},\ \bibinfo {pages} {2050} (\bibinfo {year}
  {1999})}\BibitemShut {NoStop}%
\bibitem [{\citenamefont {Shor}\ and\ \citenamefont
  {Preskill}(2000)}]{shor2000simple}%
  \BibitemOpen
  \bibfield  {author} {\bibinfo {author} {\bibfnamefont {P.~W.}\ \bibnamefont
  {Shor}}\ and\ \bibinfo {author} {\bibfnamefont {J.}~\bibnamefont
  {Preskill}},\ }\href@noop {} {\bibfield  {journal} {\bibinfo  {journal}
  {Physical review letters}\ }\textbf {\bibinfo {volume} {85}},\ \bibinfo
  {pages} {441} (\bibinfo {year} {2000})}\BibitemShut {NoStop}%
\bibitem [{\citenamefont {Scarani}\ \emph {et~al.}(2009)\citenamefont
  {Scarani}, \citenamefont {Bechmann-Pasquinucci}, \citenamefont {Cerf},
  \citenamefont {Du\ifmmode~\check{s}\else \v{s}\fi{}ek}, \citenamefont
  {L\"utkenhaus},\ and\ \citenamefont {Peev}}]{Scarani:QKDrev:2009}%
  \BibitemOpen
  \bibfield  {author} {\bibinfo {author} {\bibfnamefont {V.}~\bibnamefont
  {Scarani}}, \bibinfo {author} {\bibfnamefont {H.}~\bibnamefont
  {Bechmann-Pasquinucci}}, \bibinfo {author} {\bibfnamefont {N.~J.}\
  \bibnamefont {Cerf}}, \bibinfo {author} {\bibfnamefont {M.}~\bibnamefont
  {Du\ifmmode~\check{s}\else \v{s}\fi{}ek}}, \bibinfo {author} {\bibfnamefont
  {N.}~\bibnamefont {L\"utkenhaus}}, \ and\ \bibinfo {author} {\bibfnamefont
  {M.}~\bibnamefont {Peev}},\ }\href {\doibase 10.1103/RevModPhys.81.1301}
  {\bibfield  {journal} {\bibinfo  {journal} {\rmp}\ }\textbf {\bibinfo
  {volume} {81}},\ \bibinfo {pages} {1301} (\bibinfo {year}
  {2009})}\BibitemShut {NoStop}%
\bibitem [{\citenamefont {Renner}(2008)}]{Rennersecurity}%
  \BibitemOpen
  \bibfield  {author} {\bibinfo {author} {\bibfnamefont {R.}~\bibnamefont
  {Renner}},\ }\href@noop {} {\bibfield  {journal} {\bibinfo  {journal} {Int.\
  J.\ Quantum\ Inf.}\ }\textbf {\bibinfo {volume} {6}},\ \bibinfo {pages} {1}
  (\bibinfo {year} {2008})}\BibitemShut {NoStop}%
\bibitem [{\citenamefont {Hwang}(2003)}]{hwang2003quantum}%
  \BibitemOpen
  \bibfield  {author} {\bibinfo {author} {\bibfnamefont {W.-Y.}\ \bibnamefont
  {Hwang}},\ }\href@noop {} {\bibfield  {journal} {\bibinfo  {journal} {\prl}\
  }\textbf {\bibinfo {volume} {91}},\ \bibinfo {pages} {057901} (\bibinfo
  {year} {2003})}\BibitemShut {NoStop}%
\bibitem [{\citenamefont {Wang}(2005{\natexlab{a}})}]{wang2005beating}%
  \BibitemOpen
  \bibfield  {author} {\bibinfo {author} {\bibfnamefont {X.-B.}\ \bibnamefont
  {Wang}},\ }\href@noop {} {\bibfield  {journal} {\bibinfo  {journal} {Physical
  Review Letters}\ }\textbf {\bibinfo {volume} {94}},\ \bibinfo {pages}
  {230503} (\bibinfo {year} {2005}{\natexlab{a}})}\BibitemShut {NoStop}%
\bibitem [{\citenamefont {Lo}\ \emph {et~al.}(2005)\citenamefont {Lo},
  \citenamefont {Ma},\ and\ \citenamefont {Chen}}]{lo2005decoy}%
  \BibitemOpen
  \bibfield  {author} {\bibinfo {author} {\bibfnamefont {H.-K.}\ \bibnamefont
  {Lo}}, \bibinfo {author} {\bibfnamefont {X.}~\bibnamefont {Ma}}, \ and\
  \bibinfo {author} {\bibfnamefont {K.}~\bibnamefont {Chen}},\ }\href@noop {}
  {\bibfield  {journal} {\bibinfo  {journal} {\prl}\ }\textbf {\bibinfo
  {volume} {94}},\ \bibinfo {pages} {230504} (\bibinfo {year}
  {2005})}\BibitemShut {NoStop}%
\bibitem [{\citenamefont {Boaron}\ \emph {et~al.}(2018)\citenamefont {Boaron},
  \citenamefont {Boso}, \citenamefont {Rusca}, \citenamefont {Vulliez},
  \citenamefont {Autebert}, \citenamefont {Caloz}, \citenamefont {Perrenoud},
  \citenamefont {Gras}, \citenamefont {Bussi{\`e}res}, \citenamefont {Li} \emph
  {et~al.}}]{boaron2018secure}%
  \BibitemOpen
  \bibfield  {author} {\bibinfo {author} {\bibfnamefont {A.}~\bibnamefont
  {Boaron}}, \bibinfo {author} {\bibfnamefont {G.}~\bibnamefont {Boso}},
  \bibinfo {author} {\bibfnamefont {D.}~\bibnamefont {Rusca}}, \bibinfo
  {author} {\bibfnamefont {C.}~\bibnamefont {Vulliez}}, \bibinfo {author}
  {\bibfnamefont {C.}~\bibnamefont {Autebert}}, \bibinfo {author}
  {\bibfnamefont {M.}~\bibnamefont {Caloz}}, \bibinfo {author} {\bibfnamefont
  {M.}~\bibnamefont {Perrenoud}}, \bibinfo {author} {\bibfnamefont
  {G.}~\bibnamefont {Gras}}, \bibinfo {author} {\bibfnamefont {F.}~\bibnamefont
  {Bussi{\`e}res}}, \bibinfo {author} {\bibfnamefont {M.-J.}\ \bibnamefont
  {Li}},  \emph {et~al.},\ }\href@noop {} {\bibfield  {journal} {\bibinfo
  {journal} {Physical review letters}\ }\textbf {\bibinfo {volume} {121}},\
  \bibinfo {pages} {190502} (\bibinfo {year} {2018})}\BibitemShut {NoStop}%
\bibitem [{\citenamefont {Yin}\ \emph {et~al.}(2016)\citenamefont {Yin},
  \citenamefont {Chen}, \citenamefont {Yu}, \citenamefont {Liu}, \citenamefont
  {You}, \citenamefont {Zhou}, \citenamefont {Chen}, \citenamefont {Mao},
  \citenamefont {Huang}, \citenamefont {Zhang} \emph
  {et~al.}}]{yin2016measurement}%
  \BibitemOpen
  \bibfield  {author} {\bibinfo {author} {\bibfnamefont {H.-L.}\ \bibnamefont
  {Yin}}, \bibinfo {author} {\bibfnamefont {T.-Y.}\ \bibnamefont {Chen}},
  \bibinfo {author} {\bibfnamefont {Z.-W.}\ \bibnamefont {Yu}}, \bibinfo
  {author} {\bibfnamefont {H.}~\bibnamefont {Liu}}, \bibinfo {author}
  {\bibfnamefont {L.-X.}\ \bibnamefont {You}}, \bibinfo {author} {\bibfnamefont
  {Y.-H.}\ \bibnamefont {Zhou}}, \bibinfo {author} {\bibfnamefont {S.-J.}\
  \bibnamefont {Chen}}, \bibinfo {author} {\bibfnamefont {Y.}~\bibnamefont
  {Mao}}, \bibinfo {author} {\bibfnamefont {M.-Q.}\ \bibnamefont {Huang}},
  \bibinfo {author} {\bibfnamefont {W.-J.}\ \bibnamefont {Zhang}},  \emph
  {et~al.},\ }\href@noop {} {\bibfield  {journal} {\bibinfo  {journal}
  {Physical review letters}\ }\textbf {\bibinfo {volume} {117}},\ \bibinfo
  {pages} {190501} (\bibinfo {year} {2016})}\BibitemShut {NoStop}%
\bibitem [{\citenamefont {Wang}\ \emph {et~al.}(2012)\citenamefont {Wang},
  \citenamefont {Chen}, \citenamefont {Guo}, \citenamefont {Yin}, \citenamefont
  {Li}, \citenamefont {Zhou}, \citenamefont {Guo},\ and\ \citenamefont
  {Han}}]{wang20122}%
  \BibitemOpen
  \bibfield  {author} {\bibinfo {author} {\bibfnamefont {S.}~\bibnamefont
  {Wang}}, \bibinfo {author} {\bibfnamefont {W.}~\bibnamefont {Chen}}, \bibinfo
  {author} {\bibfnamefont {J.-F.}\ \bibnamefont {Guo}}, \bibinfo {author}
  {\bibfnamefont {Z.-Q.}\ \bibnamefont {Yin}}, \bibinfo {author} {\bibfnamefont
  {H.-W.}\ \bibnamefont {Li}}, \bibinfo {author} {\bibfnamefont
  {Z.}~\bibnamefont {Zhou}}, \bibinfo {author} {\bibfnamefont {G.-C.}\
  \bibnamefont {Guo}}, \ and\ \bibinfo {author} {\bibfnamefont {Z.-F.}\
  \bibnamefont {Han}},\ }\href@noop {} {\bibfield  {journal} {\bibinfo
  {journal} {Optics letters}\ }\textbf {\bibinfo {volume} {37}},\ \bibinfo
  {pages} {1008} (\bibinfo {year} {2012})}\BibitemShut {NoStop}%
\bibitem [{\citenamefont {Gordon}\ \emph {et~al.}(2005)\citenamefont {Gordon},
  \citenamefont {Fernandez}, \citenamefont {Buller}, \citenamefont {Rech},
  \citenamefont {Cova},\ and\ \citenamefont {Townsend}}]{gordon2005quantum}%
  \BibitemOpen
  \bibfield  {author} {\bibinfo {author} {\bibfnamefont {K.~J.}\ \bibnamefont
  {Gordon}}, \bibinfo {author} {\bibfnamefont {V.}~\bibnamefont {Fernandez}},
  \bibinfo {author} {\bibfnamefont {G.~S.}\ \bibnamefont {Buller}}, \bibinfo
  {author} {\bibfnamefont {I.}~\bibnamefont {Rech}}, \bibinfo {author}
  {\bibfnamefont {S.~D.}\ \bibnamefont {Cova}}, \ and\ \bibinfo {author}
  {\bibfnamefont {P.~D.}\ \bibnamefont {Townsend}},\ }\href@noop {} {\bibfield
  {journal} {\bibinfo  {journal} {Optics Express}\ }\textbf {\bibinfo {volume}
  {13}},\ \bibinfo {pages} {3015} (\bibinfo {year} {2005})}\BibitemShut
  {NoStop}%
\bibitem [{\citenamefont {Wang}\ \emph {et~al.}(2018)\citenamefont {Wang},
  \citenamefont {Chen}, \citenamefont {Yin}, \citenamefont {He}, \citenamefont
  {Hui}, \citenamefont {Hao}, \citenamefont {Fan-Yuan}, \citenamefont {Wang},
  \citenamefont {Zhang}, \citenamefont {Kuang} \emph
  {et~al.}}]{wang2018practical}%
  \BibitemOpen
  \bibfield  {author} {\bibinfo {author} {\bibfnamefont {S.}~\bibnamefont
  {Wang}}, \bibinfo {author} {\bibfnamefont {W.}~\bibnamefont {Chen}}, \bibinfo
  {author} {\bibfnamefont {Z.-Q.}\ \bibnamefont {Yin}}, \bibinfo {author}
  {\bibfnamefont {D.-Y.}\ \bibnamefont {He}}, \bibinfo {author} {\bibfnamefont
  {C.}~\bibnamefont {Hui}}, \bibinfo {author} {\bibfnamefont {P.-L.}\
  \bibnamefont {Hao}}, \bibinfo {author} {\bibfnamefont {G.-J.}\ \bibnamefont
  {Fan-Yuan}}, \bibinfo {author} {\bibfnamefont {C.}~\bibnamefont {Wang}},
  \bibinfo {author} {\bibfnamefont {L.-J.}\ \bibnamefont {Zhang}}, \bibinfo
  {author} {\bibfnamefont {J.}~\bibnamefont {Kuang}},  \emph {et~al.},\
  }\href@noop {} {\bibfield  {journal} {\bibinfo  {journal} {Optics letters}\
  }\textbf {\bibinfo {volume} {43}},\ \bibinfo {pages} {2030} (\bibinfo {year}
  {2018})}\BibitemShut {NoStop}%
\bibitem [{\citenamefont {Takesue}\ \emph {et~al.}(2007)\citenamefont
  {Takesue}, \citenamefont {Nam}, \citenamefont {Zhang}, \citenamefont
  {Hadfield}, \citenamefont {Honjo}, \citenamefont {Tamaki},\ and\
  \citenamefont {Yamamoto}}]{takesue2007quantum}%
  \BibitemOpen
  \bibfield  {author} {\bibinfo {author} {\bibfnamefont {H.}~\bibnamefont
  {Takesue}}, \bibinfo {author} {\bibfnamefont {S.~W.}\ \bibnamefont {Nam}},
  \bibinfo {author} {\bibfnamefont {Q.}~\bibnamefont {Zhang}}, \bibinfo
  {author} {\bibfnamefont {R.~H.}\ \bibnamefont {Hadfield}}, \bibinfo {author}
  {\bibfnamefont {T.}~\bibnamefont {Honjo}}, \bibinfo {author} {\bibfnamefont
  {K.}~\bibnamefont {Tamaki}}, \ and\ \bibinfo {author} {\bibfnamefont
  {Y.}~\bibnamefont {Yamamoto}},\ }\href@noop {} {\bibfield  {journal}
  {\bibinfo  {journal} {Nature photonics}\ }\textbf {\bibinfo {volume} {1}},\
  \bibinfo {pages} {343} (\bibinfo {year} {2007})}\BibitemShut {NoStop}%
\bibitem [{\citenamefont {Lucamarini}\ \emph {et~al.}(2018)\citenamefont
  {Lucamarini}, \citenamefont {Yuan}, \citenamefont {Dynes},\ and\
  \citenamefont {Shields}}]{tfqkd}%
  \BibitemOpen
  \bibfield  {author} {\bibinfo {author} {\bibfnamefont {M.}~\bibnamefont
  {Lucamarini}}, \bibinfo {author} {\bibfnamefont {Z.}~\bibnamefont {Yuan}},
  \bibinfo {author} {\bibfnamefont {J.}~\bibnamefont {Dynes}}, \ and\ \bibinfo
  {author} {\bibfnamefont {A.}~\bibnamefont {Shields}},\ }\href@noop {}
  {\bibfield  {journal} {\bibinfo  {journal} {Nature}\ }\textbf {\bibinfo
  {volume} {557}},\ \bibinfo {pages} {400} (\bibinfo {year}
  {2018})}\BibitemShut {NoStop}%
\bibitem [{\citenamefont {Lo}\ \emph {et~al.}(2012)\citenamefont {Lo},
  \citenamefont {Curty},\ and\ \citenamefont {Qi}}]{lo2012measurement}%
  \BibitemOpen
  \bibfield  {author} {\bibinfo {author} {\bibfnamefont {H.-K.}\ \bibnamefont
  {Lo}}, \bibinfo {author} {\bibfnamefont {M.}~\bibnamefont {Curty}}, \ and\
  \bibinfo {author} {\bibfnamefont {B.}~\bibnamefont {Qi}},\ }\href@noop {}
  {\bibfield  {journal} {\bibinfo  {journal} {Physical review letters}\
  }\textbf {\bibinfo {volume} {108}},\ \bibinfo {pages} {130503} (\bibinfo
  {year} {2012})}\BibitemShut {NoStop}%
\bibitem [{\citenamefont {Ac{\'\i}n}\ \emph {et~al.}(2007)\citenamefont
  {Ac{\'\i}n}, \citenamefont {Brunner}, \citenamefont {Gisin}, \citenamefont
  {Massar}, \citenamefont {Pironio},\ and\ \citenamefont
  {Scarani}}]{acin2007device}%
  \BibitemOpen
  \bibfield  {author} {\bibinfo {author} {\bibfnamefont {A.}~\bibnamefont
  {Ac{\'\i}n}}, \bibinfo {author} {\bibfnamefont {N.}~\bibnamefont {Brunner}},
  \bibinfo {author} {\bibfnamefont {N.}~\bibnamefont {Gisin}}, \bibinfo
  {author} {\bibfnamefont {S.}~\bibnamefont {Massar}}, \bibinfo {author}
  {\bibfnamefont {S.}~\bibnamefont {Pironio}}, \ and\ \bibinfo {author}
  {\bibfnamefont {V.}~\bibnamefont {Scarani}},\ }\href@noop {} {\bibfield
  {journal} {\bibinfo  {journal} {Physical Review Letters}\ }\textbf {\bibinfo
  {volume} {98}},\ \bibinfo {pages} {230501} (\bibinfo {year}
  {2007})}\BibitemShut {NoStop}%
\bibitem [{\citenamefont {Sasaki}\ \emph {et~al.}(2014)\citenamefont {Sasaki},
  \citenamefont {Yamamoto},\ and\ \citenamefont
  {Koashi}}]{sasaki2014practical}%
  \BibitemOpen
  \bibfield  {author} {\bibinfo {author} {\bibfnamefont {T.}~\bibnamefont
  {Sasaki}}, \bibinfo {author} {\bibfnamefont {Y.}~\bibnamefont {Yamamoto}}, \
  and\ \bibinfo {author} {\bibfnamefont {M.}~\bibnamefont {Koashi}},\
  }\href@noop {} {\bibfield  {journal} {\bibinfo  {journal} {Nature}\ }\textbf
  {\bibinfo {volume} {509}},\ \bibinfo {pages} {475} (\bibinfo {year}
  {2014})}\BibitemShut {NoStop}%
\bibitem [{\citenamefont {Laing}\ \emph {et~al.}(2010)\citenamefont {Laing},
  \citenamefont {Scarani}, \citenamefont {Rarity},\ and\ \citenamefont
  {O’Brien}}]{laing2010reference}%
  \BibitemOpen
  \bibfield  {author} {\bibinfo {author} {\bibfnamefont {A.}~\bibnamefont
  {Laing}}, \bibinfo {author} {\bibfnamefont {V.}~\bibnamefont {Scarani}},
  \bibinfo {author} {\bibfnamefont {J.~G.}\ \bibnamefont {Rarity}}, \ and\
  \bibinfo {author} {\bibfnamefont {J.~L.}\ \bibnamefont {O’Brien}},\
  }\href@noop {} {\bibfield  {journal} {\bibinfo  {journal} {Physical Review
  A}\ }\textbf {\bibinfo {volume} {82}},\ \bibinfo {pages} {012304} (\bibinfo
  {year} {2010})}\BibitemShut {NoStop}%
\bibitem [{\citenamefont {Fung}\ \emph {et~al.}(2007)\citenamefont {Fung},
  \citenamefont {Qi}, \citenamefont {Tamaki},\ and\ \citenamefont
  {Lo}}]{fung2007phase}%
  \BibitemOpen
  \bibfield  {author} {\bibinfo {author} {\bibfnamefont {C.-H.~F.}\
  \bibnamefont {Fung}}, \bibinfo {author} {\bibfnamefont {B.}~\bibnamefont
  {Qi}}, \bibinfo {author} {\bibfnamefont {K.}~\bibnamefont {Tamaki}}, \ and\
  \bibinfo {author} {\bibfnamefont {H.-K.}\ \bibnamefont {Lo}},\ }\href@noop {}
  {\bibfield  {journal} {\bibinfo  {journal} {Physical Review A}\ }\textbf
  {\bibinfo {volume} {75}},\ \bibinfo {pages} {032314} (\bibinfo {year}
  {2007})}\BibitemShut {NoStop}%
\bibitem [{\citenamefont {Zhao}\ \emph {et~al.}(2008)\citenamefont {Zhao},
  \citenamefont {Fung}, \citenamefont {Qi}, \citenamefont {Chen},\ and\
  \citenamefont {Lo}}]{zhao2008quantum}%
  \BibitemOpen
  \bibfield  {author} {\bibinfo {author} {\bibfnamefont {Y.}~\bibnamefont
  {Zhao}}, \bibinfo {author} {\bibfnamefont {C.-H.~F.}\ \bibnamefont {Fung}},
  \bibinfo {author} {\bibfnamefont {B.}~\bibnamefont {Qi}}, \bibinfo {author}
  {\bibfnamefont {C.}~\bibnamefont {Chen}}, \ and\ \bibinfo {author}
  {\bibfnamefont {H.-K.}\ \bibnamefont {Lo}},\ }\href@noop {} {\bibfield
  {journal} {\bibinfo  {journal} {Physical Review A}\ }\textbf {\bibinfo
  {volume} {78}},\ \bibinfo {pages} {042333} (\bibinfo {year}
  {2008})}\BibitemShut {NoStop}%
\bibitem [{\citenamefont {Gerhardt}\ \emph {et~al.}(2011)\citenamefont
  {Gerhardt}, \citenamefont {Liu}, \citenamefont {Lamas-Linares}, \citenamefont
  {Skaar}, \citenamefont {Kurtsiefer},\ and\ \citenamefont
  {Makarov}}]{gerhardt2011full}%
  \BibitemOpen
  \bibfield  {author} {\bibinfo {author} {\bibfnamefont {I.}~\bibnamefont
  {Gerhardt}}, \bibinfo {author} {\bibfnamefont {Q.}~\bibnamefont {Liu}},
  \bibinfo {author} {\bibfnamefont {A.}~\bibnamefont {Lamas-Linares}}, \bibinfo
  {author} {\bibfnamefont {J.}~\bibnamefont {Skaar}}, \bibinfo {author}
  {\bibfnamefont {C.}~\bibnamefont {Kurtsiefer}}, \ and\ \bibinfo {author}
  {\bibfnamefont {V.}~\bibnamefont {Makarov}},\ }\href@noop {} {\bibfield
  {journal} {\bibinfo  {journal} {Nature communications}\ }\textbf {\bibinfo
  {volume} {2}},\ \bibinfo {pages} {349} (\bibinfo {year} {2011})}\BibitemShut
  {NoStop}%
\bibitem [{\citenamefont {Makarov}(2009)}]{makarov2009controlling}%
  \BibitemOpen
  \bibfield  {author} {\bibinfo {author} {\bibfnamefont {V.}~\bibnamefont
  {Makarov}},\ }\href@noop {} {\bibfield  {journal} {\bibinfo  {journal} {New
  Journal of Physics}\ }\textbf {\bibinfo {volume} {11}},\ \bibinfo {pages}
  {065003} (\bibinfo {year} {2009})}\BibitemShut {NoStop}%
\bibitem [{\citenamefont {Liu}\ \emph {et~al.}(2013)\citenamefont {Liu},
  \citenamefont {Chen}, \citenamefont {Wang}, \citenamefont {Liang},
  \citenamefont {Shentu}, \citenamefont {Wang}, \citenamefont {Cui},
  \citenamefont {Yin}, \citenamefont {Liu}, \citenamefont {Li} \emph
  {et~al.}}]{liu2013experimental}%
  \BibitemOpen
  \bibfield  {author} {\bibinfo {author} {\bibfnamefont {Y.}~\bibnamefont
  {Liu}}, \bibinfo {author} {\bibfnamefont {T.-Y.}\ \bibnamefont {Chen}},
  \bibinfo {author} {\bibfnamefont {L.-J.}\ \bibnamefont {Wang}}, \bibinfo
  {author} {\bibfnamefont {H.}~\bibnamefont {Liang}}, \bibinfo {author}
  {\bibfnamefont {G.-L.}\ \bibnamefont {Shentu}}, \bibinfo {author}
  {\bibfnamefont {J.}~\bibnamefont {Wang}}, \bibinfo {author} {\bibfnamefont
  {K.}~\bibnamefont {Cui}}, \bibinfo {author} {\bibfnamefont {H.-L.}\
  \bibnamefont {Yin}}, \bibinfo {author} {\bibfnamefont {N.-L.}\ \bibnamefont
  {Liu}}, \bibinfo {author} {\bibfnamefont {L.}~\bibnamefont {Li}},  \emph
  {et~al.},\ }\href@noop {} {\bibfield  {journal} {\bibinfo  {journal}
  {Physical review letters}\ }\textbf {\bibinfo {volume} {111}},\ \bibinfo
  {pages} {130502} (\bibinfo {year} {2013})}\BibitemShut {NoStop}%
\bibitem [{\citenamefont {Da~Silva}\ \emph {et~al.}(2013)\citenamefont
  {Da~Silva}, \citenamefont {Vitoreti}, \citenamefont {Xavier}, \citenamefont
  {Do~Amaral}, \citenamefont {Temporao},\ and\ \citenamefont {Von
  Der~Weid}}]{da2013proof}%
  \BibitemOpen
  \bibfield  {author} {\bibinfo {author} {\bibfnamefont {T.~F.}\ \bibnamefont
  {Da~Silva}}, \bibinfo {author} {\bibfnamefont {D.}~\bibnamefont {Vitoreti}},
  \bibinfo {author} {\bibfnamefont {G.}~\bibnamefont {Xavier}}, \bibinfo
  {author} {\bibfnamefont {G.}~\bibnamefont {Do~Amaral}}, \bibinfo {author}
  {\bibfnamefont {G.}~\bibnamefont {Temporao}}, \ and\ \bibinfo {author}
  {\bibfnamefont {J.}~\bibnamefont {Von Der~Weid}},\ }\href@noop {} {\bibfield
  {journal} {\bibinfo  {journal} {Physical Review A}\ }\textbf {\bibinfo
  {volume} {88}},\ \bibinfo {pages} {052303} (\bibinfo {year}
  {2013})}\BibitemShut {NoStop}%
\bibitem [{\citenamefont {Rubenok}\ \emph {et~al.}(2013)\citenamefont
  {Rubenok}, \citenamefont {Slater}, \citenamefont {Chan}, \citenamefont
  {Lucio-Martinez},\ and\ \citenamefont {Tittel}}]{rubenok2013real}%
  \BibitemOpen
  \bibfield  {author} {\bibinfo {author} {\bibfnamefont {A.}~\bibnamefont
  {Rubenok}}, \bibinfo {author} {\bibfnamefont {J.~A.}\ \bibnamefont {Slater}},
  \bibinfo {author} {\bibfnamefont {P.}~\bibnamefont {Chan}}, \bibinfo {author}
  {\bibfnamefont {I.}~\bibnamefont {Lucio-Martinez}}, \ and\ \bibinfo {author}
  {\bibfnamefont {W.}~\bibnamefont {Tittel}},\ }\href@noop {} {\bibfield
  {journal} {\bibinfo  {journal} {Physical review letters}\ }\textbf {\bibinfo
  {volume} {111}},\ \bibinfo {pages} {130501} (\bibinfo {year}
  {2013})}\BibitemShut {NoStop}%
\bibitem [{\citenamefont {Comandar}\ \emph {et~al.}(2016)\citenamefont
  {Comandar}, \citenamefont {Lucamarini}, \citenamefont {Fr{\"o}hlich},
  \citenamefont {Dynes}, \citenamefont {Sharpe}, \citenamefont {Tam},
  \citenamefont {Yuan}, \citenamefont {Penty},\ and\ \citenamefont
  {Shields}}]{comandar2016quantum}%
  \BibitemOpen
  \bibfield  {author} {\bibinfo {author} {\bibfnamefont {L.}~\bibnamefont
  {Comandar}}, \bibinfo {author} {\bibfnamefont {M.}~\bibnamefont
  {Lucamarini}}, \bibinfo {author} {\bibfnamefont {B.}~\bibnamefont
  {Fr{\"o}hlich}}, \bibinfo {author} {\bibfnamefont {J.}~\bibnamefont {Dynes}},
  \bibinfo {author} {\bibfnamefont {A.}~\bibnamefont {Sharpe}}, \bibinfo
  {author} {\bibfnamefont {S.-B.}\ \bibnamefont {Tam}}, \bibinfo {author}
  {\bibfnamefont {Z.}~\bibnamefont {Yuan}}, \bibinfo {author} {\bibfnamefont
  {R.}~\bibnamefont {Penty}}, \ and\ \bibinfo {author} {\bibfnamefont
  {A.}~\bibnamefont {Shields}},\ }\href@noop {} {\bibfield  {journal} {\bibinfo
   {journal} {Nature Photonics}\ }\textbf {\bibinfo {volume} {10}},\ \bibinfo
  {pages} {312} (\bibinfo {year} {2016})}\BibitemShut {NoStop}%
\bibitem [{\citenamefont {Spedalieri}(2006)}]{spedalieri2006quantum}%
  \BibitemOpen
  \bibfield  {author} {\bibinfo {author} {\bibfnamefont {F.~M.}\ \bibnamefont
  {Spedalieri}},\ }\href@noop {} {\bibfield  {journal} {\bibinfo  {journal}
  {Optics communications}\ }\textbf {\bibinfo {volume} {260}},\ \bibinfo
  {pages} {340} (\bibinfo {year} {2006})}\BibitemShut {NoStop}%
\bibitem [{\citenamefont {Yin}\ \emph {et~al.}(2014)\citenamefont {Yin},
  \citenamefont {Wang}, \citenamefont {Chen}, \citenamefont {Li}, \citenamefont
  {Guo},\ and\ \citenamefont {Han}}]{yin2014reference}%
  \BibitemOpen
  \bibfield  {author} {\bibinfo {author} {\bibfnamefont {Z.-Q.}\ \bibnamefont
  {Yin}}, \bibinfo {author} {\bibfnamefont {S.}~\bibnamefont {Wang}}, \bibinfo
  {author} {\bibfnamefont {W.}~\bibnamefont {Chen}}, \bibinfo {author}
  {\bibfnamefont {H.-W.}\ \bibnamefont {Li}}, \bibinfo {author} {\bibfnamefont
  {G.-C.}\ \bibnamefont {Guo}}, \ and\ \bibinfo {author} {\bibfnamefont
  {Z.-F.}\ \bibnamefont {Han}},\ }\href@noop {} {\bibfield  {journal} {\bibinfo
   {journal} {Quantum information processing}\ }\textbf {\bibinfo {volume}
  {13}},\ \bibinfo {pages} {1237} (\bibinfo {year} {2014})}\BibitemShut
  {NoStop}%
\bibitem [{\citenamefont {Wang}\ \emph {et~al.}(2015)\citenamefont {Wang},
  \citenamefont {Song}, \citenamefont {Yin}, \citenamefont {Wang},
  \citenamefont {Chen}, \citenamefont {Zhang}, \citenamefont {Guo},\ and\
  \citenamefont {Han}}]{wang2015phase}%
  \BibitemOpen
  \bibfield  {author} {\bibinfo {author} {\bibfnamefont {C.}~\bibnamefont
  {Wang}}, \bibinfo {author} {\bibfnamefont {X.-T.}\ \bibnamefont {Song}},
  \bibinfo {author} {\bibfnamefont {Z.-Q.}\ \bibnamefont {Yin}}, \bibinfo
  {author} {\bibfnamefont {S.}~\bibnamefont {Wang}}, \bibinfo {author}
  {\bibfnamefont {W.}~\bibnamefont {Chen}}, \bibinfo {author} {\bibfnamefont
  {C.-M.}\ \bibnamefont {Zhang}}, \bibinfo {author} {\bibfnamefont {G.-C.}\
  \bibnamefont {Guo}}, \ and\ \bibinfo {author} {\bibfnamefont {Z.-F.}\
  \bibnamefont {Han}},\ }\href@noop {} {\bibfield  {journal} {\bibinfo
  {journal} {Physical review letters}\ }\textbf {\bibinfo {volume} {115}},\
  \bibinfo {pages} {160502} (\bibinfo {year} {2015})}\BibitemShut {NoStop}%
\bibitem [{\citenamefont {Xu}\ \emph {et~al.}(2014)\citenamefont {Xu},
  \citenamefont {Xu},\ and\ \citenamefont {Lo}}]{xu2014protocol}%
  \BibitemOpen
  \bibfield  {author} {\bibinfo {author} {\bibfnamefont {F.}~\bibnamefont
  {Xu}}, \bibinfo {author} {\bibfnamefont {H.}~\bibnamefont {Xu}}, \ and\
  \bibinfo {author} {\bibfnamefont {H.-K.}\ \bibnamefont {Lo}},\ }\href@noop {}
  {\bibfield  {journal} {\bibinfo  {journal} {Physical Review A}\ }\textbf
  {\bibinfo {volume} {89}},\ \bibinfo {pages} {052333} (\bibinfo {year}
  {2014})}\BibitemShut {NoStop}%
\bibitem [{\citenamefont {Ma}\ and\ \citenamefont
  {Razavi}(2012{\natexlab{a}})}]{ma2012statistical}%
  \BibitemOpen
  \bibfield  {author} {\bibinfo {author} {\bibfnamefont {C.-H. F.~F.}\
  \bibnamefont {Ma}, \bibfnamefont {Xiongfeng}}\ and\ \bibinfo {author}
  {\bibfnamefont {M.}~\bibnamefont {Razavi}},\ }\href@noop {} {\bibfield
  {journal} {\bibinfo  {journal} {Physical Review A}\ }\textbf {\bibinfo
  {volume} {86}},\ \bibinfo {pages} {052305} (\bibinfo {year}
  {2012}{\natexlab{a}})}\BibitemShut {NoStop}%
\bibitem [{\citenamefont {Wang}(2013)}]{wang2013three}%
  \BibitemOpen
  \bibfield  {author} {\bibinfo {author} {\bibfnamefont {X.-B.}\ \bibnamefont
  {Wang}},\ }\href@noop {} {\bibfield  {journal} {\bibinfo  {journal} {Physical
  Review A}\ }\textbf {\bibinfo {volume} {87}},\ \bibinfo {pages} {012320}
  (\bibinfo {year} {2013})}\BibitemShut {NoStop}%
\bibitem [{\citenamefont {Yu}\ \emph {et~al.}(2015)\citenamefont {Yu},
  \citenamefont {Zhou},\ and\ \citenamefont {Wang}}]{yu2015statistical}%
  \BibitemOpen
  \bibfield  {author} {\bibinfo {author} {\bibfnamefont {Z.-W.}\ \bibnamefont
  {Yu}}, \bibinfo {author} {\bibfnamefont {Y.-H.}\ \bibnamefont {Zhou}}, \ and\
  \bibinfo {author} {\bibfnamefont {X.-B.}\ \bibnamefont {Wang}},\ }\href@noop
  {} {\bibfield  {journal} {\bibinfo  {journal} {Physical Review A}\ }\textbf
  {\bibinfo {volume} {91}},\ \bibinfo {pages} {032318} (\bibinfo {year}
  {2015})}\BibitemShut {NoStop}%
\bibitem [{\citenamefont {Zhou}\ \emph {et~al.}(2016)\citenamefont {Zhou},
  \citenamefont {Yu},\ and\ \citenamefont {Wang}}]{zhou2016making}%
  \BibitemOpen
  \bibfield  {author} {\bibinfo {author} {\bibfnamefont {Y.-H.}\ \bibnamefont
  {Zhou}}, \bibinfo {author} {\bibfnamefont {Z.-W.}\ \bibnamefont {Yu}}, \ and\
  \bibinfo {author} {\bibfnamefont {X.-B.}\ \bibnamefont {Wang}},\ }\href@noop
  {} {\bibfield  {journal} {\bibinfo  {journal} {Physical Review A}\ }\textbf
  {\bibinfo {volume} {93}},\ \bibinfo {pages} {042324} (\bibinfo {year}
  {2016})}\BibitemShut {NoStop}%
\bibitem [{\citenamefont {Curty}\ \emph {et~al.}(2014)\citenamefont {Curty},
  \citenamefont {Xu}, \citenamefont {Cui}, \citenamefont {Lim}, \citenamefont
  {Tamaki},\ and\ \citenamefont {Lo}}]{curty2014finite}%
  \BibitemOpen
  \bibfield  {author} {\bibinfo {author} {\bibfnamefont {M.}~\bibnamefont
  {Curty}}, \bibinfo {author} {\bibfnamefont {F.}~\bibnamefont {Xu}}, \bibinfo
  {author} {\bibfnamefont {W.}~\bibnamefont {Cui}}, \bibinfo {author}
  {\bibfnamefont {C.~C.~W.}\ \bibnamefont {Lim}}, \bibinfo {author}
  {\bibfnamefont {K.}~\bibnamefont {Tamaki}}, \ and\ \bibinfo {author}
  {\bibfnamefont {H.-K.}\ \bibnamefont {Lo}},\ }\href@noop {} {\bibfield
  {journal} {\bibinfo  {journal} {Nature communications}\ }\textbf {\bibinfo
  {volume} {5}},\ \bibinfo {pages} {3732} (\bibinfo {year} {2014})}\BibitemShut
  {NoStop}%
\bibitem [{\citenamefont {Sheridan}\ \emph {et~al.}(2010)\citenamefont
  {Sheridan}, \citenamefont {Le},\ and\ \citenamefont
  {Scarani}}]{sheridan2010finite}%
  \BibitemOpen
  \bibfield  {author} {\bibinfo {author} {\bibfnamefont {L.}~\bibnamefont
  {Sheridan}}, \bibinfo {author} {\bibfnamefont {T.~P.}\ \bibnamefont {Le}}, \
  and\ \bibinfo {author} {\bibfnamefont {V.}~\bibnamefont {Scarani}},\
  }\href@noop {} {\bibfield  {journal} {\bibinfo  {journal} {New Journal of
  Physics}\ }\textbf {\bibinfo {volume} {12}},\ \bibinfo {pages} {123019}
  (\bibinfo {year} {2010})}\BibitemShut {NoStop}%
\bibitem [{\citenamefont {Wang}\ \emph {et~al.}(2017)\citenamefont {Wang},
  \citenamefont {Yin}, \citenamefont {Wang}, \citenamefont {Chen},
  \citenamefont {Guo},\ and\ \citenamefont {Han}}]{wang2017measurement}%
  \BibitemOpen
  \bibfield  {author} {\bibinfo {author} {\bibfnamefont {C.}~\bibnamefont
  {Wang}}, \bibinfo {author} {\bibfnamefont {Z.-Q.}\ \bibnamefont {Yin}},
  \bibinfo {author} {\bibfnamefont {S.}~\bibnamefont {Wang}}, \bibinfo {author}
  {\bibfnamefont {W.}~\bibnamefont {Chen}}, \bibinfo {author} {\bibfnamefont
  {G.-C.}\ \bibnamefont {Guo}}, \ and\ \bibinfo {author} {\bibfnamefont
  {Z.-F.}\ \bibnamefont {Han}},\ }\href@noop {} {\bibfield  {journal} {\bibinfo
   {journal} {Optica}\ }\textbf {\bibinfo {volume} {4}},\ \bibinfo {pages}
  {1016} (\bibinfo {year} {2017})}\BibitemShut {NoStop}%
\bibitem [{\citenamefont {Liu}\ \emph {et~al.}(2018)\citenamefont {Liu},
  \citenamefont {Wang}, \citenamefont {Ma},\ and\ \citenamefont
  {Sun}}]{liu2018polarization}%
  \BibitemOpen
  \bibfield  {author} {\bibinfo {author} {\bibfnamefont {H.}~\bibnamefont
  {Liu}}, \bibinfo {author} {\bibfnamefont {J.}~\bibnamefont {Wang}}, \bibinfo
  {author} {\bibfnamefont {H.}~\bibnamefont {Ma}}, \ and\ \bibinfo {author}
  {\bibfnamefont {S.}~\bibnamefont {Sun}},\ }\href@noop {} {\bibfield
  {journal} {\bibinfo  {journal} {Optica}\ }\textbf {\bibinfo {volume} {5}},\
  \bibinfo {pages} {902} (\bibinfo {year} {2018})}\BibitemShut {NoStop}%
\bibitem [{\citenamefont {Zhang}\ \emph {et~al.}(2017)\citenamefont {Zhang},
  \citenamefont {Zhu},\ and\ \citenamefont {Wang}}]{zhang2017practical}%
  \BibitemOpen
  \bibfield  {author} {\bibinfo {author} {\bibfnamefont {C.-M.}\ \bibnamefont
  {Zhang}}, \bibinfo {author} {\bibfnamefont {J.-R.}\ \bibnamefont {Zhu}}, \
  and\ \bibinfo {author} {\bibfnamefont {Q.}~\bibnamefont {Wang}},\ }\href@noop
  {} {\bibfield  {journal} {\bibinfo  {journal} {Physical Review A}\ }\textbf
  {\bibinfo {volume} {95}},\ \bibinfo {pages} {032309} (\bibinfo {year}
  {2017})}\BibitemShut {NoStop}%
\bibitem [{\citenamefont {Inamori}\ \emph {et~al.}(2007)\citenamefont
  {Inamori}, \citenamefont {L{\"u}tkenhaus},\ and\ \citenamefont
  {Mayers}}]{inamori2007unconditional}%
  \BibitemOpen
  \bibfield  {author} {\bibinfo {author} {\bibfnamefont {H.}~\bibnamefont
  {Inamori}}, \bibinfo {author} {\bibfnamefont {N.}~\bibnamefont
  {L{\"u}tkenhaus}}, \ and\ \bibinfo {author} {\bibfnamefont {D.}~\bibnamefont
  {Mayers}},\ }\href@noop {} {\bibfield  {journal} {\bibinfo  {journal} {The
  European Physical Journal D}\ }\textbf {\bibinfo {volume} {41}},\ \bibinfo
  {pages} {599} (\bibinfo {year} {2007})}\BibitemShut {NoStop}%
\bibitem [{\citenamefont {Gottesman}\ \emph {et~al.}(2004)\citenamefont
  {Gottesman}, \citenamefont {Lo}, \citenamefont {Lutkenhaus},\ and\
  \citenamefont {Preskill}}]{gottesman2004security}%
  \BibitemOpen
  \bibfield  {author} {\bibinfo {author} {\bibfnamefont {D.}~\bibnamefont
  {Gottesman}}, \bibinfo {author} {\bibfnamefont {H.-K.}\ \bibnamefont {Lo}},
  \bibinfo {author} {\bibfnamefont {N.}~\bibnamefont {Lutkenhaus}}, \ and\
  \bibinfo {author} {\bibfnamefont {J.}~\bibnamefont {Preskill}},\ }in\
  \href@noop {} {\emph {\bibinfo {booktitle} {International Symposium
  onInformation Theory, 2004. ISIT 2004. Proceedings.}}}\ (\bibinfo
  {organization} {IEEE},\ \bibinfo {year} {2004})\ p.\ \bibinfo {pages}
  {136}\BibitemShut {NoStop}%
\bibitem [{\citenamefont {Shi}\ and\ \citenamefont
  {Eberhart}(1999)}]{shi1999empirical}%
  \BibitemOpen
  \bibfield  {author} {\bibinfo {author} {\bibfnamefont {Y.}~\bibnamefont
  {Shi}}\ and\ \bibinfo {author} {\bibfnamefont {R.~C.}\ \bibnamefont
  {Eberhart}},\ }in\ \href@noop {} {\emph {\bibinfo {booktitle} {Proceedings of
  the 1999 Congress on Evolutionary Computation-CEC99 (Cat. No. 99TH8406)}}},\
  Vol.~\bibinfo {volume} {3}\ (\bibinfo {organization} {IEEE},\ \bibinfo {year}
  {1999})\ pp.\ \bibinfo {pages} {1945--1950}\BibitemShut {NoStop}%
\bibitem [{\citenamefont {Thew}\ \emph {et~al.}(2006)\citenamefont {Thew},
  \citenamefont {Tanzilli}, \citenamefont {Krainer}, \citenamefont {Zeller},
  \citenamefont {Rochas}, \citenamefont {Rech}, \citenamefont {Cova},
  \citenamefont {Zbinden},\ and\ \citenamefont {Gisin}}]{thew2006low}%
  \BibitemOpen
  \bibfield  {author} {\bibinfo {author} {\bibfnamefont {R.~T.}\ \bibnamefont
  {Thew}}, \bibinfo {author} {\bibfnamefont {S.}~\bibnamefont {Tanzilli}},
  \bibinfo {author} {\bibfnamefont {L.}~\bibnamefont {Krainer}}, \bibinfo
  {author} {\bibfnamefont {S.~C.}\ \bibnamefont {Zeller}}, \bibinfo {author}
  {\bibfnamefont {A.}~\bibnamefont {Rochas}}, \bibinfo {author} {\bibfnamefont
  {I.}~\bibnamefont {Rech}}, \bibinfo {author} {\bibfnamefont {S.}~\bibnamefont
  {Cova}}, \bibinfo {author} {\bibfnamefont {H.}~\bibnamefont {Zbinden}}, \
  and\ \bibinfo {author} {\bibfnamefont {N.}~\bibnamefont {Gisin}},\
  }\href@noop {} {\bibfield  {journal} {\bibinfo  {journal} {New Journal of
  Physics}\ }\textbf {\bibinfo {volume} {8}},\ \bibinfo {pages} {32} (\bibinfo
  {year} {2006})}\BibitemShut {NoStop}%
\bibitem [{\citenamefont {Ma}\ and\ \citenamefont
  {Razavi}(2012{\natexlab{b}})}]{ma2012alternative}%
  \BibitemOpen
  \bibfield  {author} {\bibinfo {author} {\bibfnamefont {X.}~\bibnamefont
  {Ma}}\ and\ \bibinfo {author} {\bibfnamefont {M.}~\bibnamefont {Razavi}},\
  }\href@noop {} {\bibfield  {journal} {\bibinfo  {journal} {Physical Review
  A}\ }\textbf {\bibinfo {volume} {86}},\ \bibinfo {pages} {062319} (\bibinfo
  {year} {2012}{\natexlab{b}})}\BibitemShut {NoStop}%
\bibitem [{\citenamefont {Tang}\ \emph {et~al.}(2016)\citenamefont {Tang},
  \citenamefont {Yin}, \citenamefont {Zhao}, \citenamefont {Liu}, \citenamefont
  {Sun}, \citenamefont {Huang}, \citenamefont {Zhang}, \citenamefont {Chen},
  \citenamefont {Zhang}, \citenamefont {You} \emph
  {et~al.}}]{tang2016measurement}%
  \BibitemOpen
  \bibfield  {author} {\bibinfo {author} {\bibfnamefont {Y.-L.}\ \bibnamefont
  {Tang}}, \bibinfo {author} {\bibfnamefont {H.-L.}\ \bibnamefont {Yin}},
  \bibinfo {author} {\bibfnamefont {Q.}~\bibnamefont {Zhao}}, \bibinfo {author}
  {\bibfnamefont {H.}~\bibnamefont {Liu}}, \bibinfo {author} {\bibfnamefont
  {X.-X.}\ \bibnamefont {Sun}}, \bibinfo {author} {\bibfnamefont {M.-Q.}\
  \bibnamefont {Huang}}, \bibinfo {author} {\bibfnamefont {W.-J.}\ \bibnamefont
  {Zhang}}, \bibinfo {author} {\bibfnamefont {S.-J.}\ \bibnamefont {Chen}},
  \bibinfo {author} {\bibfnamefont {L.}~\bibnamefont {Zhang}}, \bibinfo
  {author} {\bibfnamefont {L.-X.}\ \bibnamefont {You}},  \emph {et~al.},\
  }\href@noop {} {\bibfield  {journal} {\bibinfo  {journal} {Physical Review
  X}\ }\textbf {\bibinfo {volume} {6}},\ \bibinfo {pages} {011024} (\bibinfo
  {year} {2016})}\BibitemShut {NoStop}%
\bibitem [{\citenamefont {Tang}\ \emph {et~al.}(2014)\citenamefont {Tang},
  \citenamefont {Yin}, \citenamefont {Chen}, \citenamefont {Liu}, \citenamefont
  {Zhang}, \citenamefont {Jiang}, \citenamefont {Zhang}, \citenamefont {Wang},
  \citenamefont {You}, \citenamefont {Guan} \emph
  {et~al.}}]{tang2014measurement}%
  \BibitemOpen
  \bibfield  {author} {\bibinfo {author} {\bibfnamefont {Y.-L.}\ \bibnamefont
  {Tang}}, \bibinfo {author} {\bibfnamefont {H.-L.}\ \bibnamefont {Yin}},
  \bibinfo {author} {\bibfnamefont {S.-J.}\ \bibnamefont {Chen}}, \bibinfo
  {author} {\bibfnamefont {Y.}~\bibnamefont {Liu}}, \bibinfo {author}
  {\bibfnamefont {W.-J.}\ \bibnamefont {Zhang}}, \bibinfo {author}
  {\bibfnamefont {X.}~\bibnamefont {Jiang}}, \bibinfo {author} {\bibfnamefont
  {L.}~\bibnamefont {Zhang}}, \bibinfo {author} {\bibfnamefont
  {J.}~\bibnamefont {Wang}}, \bibinfo {author} {\bibfnamefont {L.-X.}\
  \bibnamefont {You}}, \bibinfo {author} {\bibfnamefont {J.-Y.}\ \bibnamefont
  {Guan}},  \emph {et~al.},\ }\href@noop {} {\bibfield  {journal} {\bibinfo
  {journal} {Physical review letters}\ }\textbf {\bibinfo {volume} {113}},\
  \bibinfo {pages} {190501} (\bibinfo {year} {2014})}\BibitemShut {NoStop}%
\bibitem [{\citenamefont {Wang}(2005{\natexlab{b}})}]{wang2005decoy}%
  \BibitemOpen
  \bibfield  {author} {\bibinfo {author} {\bibfnamefont {X.-B.}\ \bibnamefont
  {Wang}},\ }\href@noop {} {\bibfield  {journal} {\bibinfo  {journal} {\pra}\
  }\textbf {\bibinfo {volume} {72}},\ \bibinfo {pages} {012322} (\bibinfo
  {year} {2005}{\natexlab{b}})}\BibitemShut {NoStop}%
\bibitem [{\citenamefont {Christandl}\ \emph {et~al.}(2009)\citenamefont
  {Christandl}, \citenamefont {K{\"o}nig},\ and\ \citenamefont
  {Renner}}]{christandl2009postselection}%
  \BibitemOpen
  \bibfield  {author} {\bibinfo {author} {\bibfnamefont {M.}~\bibnamefont
  {Christandl}}, \bibinfo {author} {\bibfnamefont {R.}~\bibnamefont
  {K{\"o}nig}}, \ and\ \bibinfo {author} {\bibfnamefont {R.}~\bibnamefont
  {Renner}},\ }\href@noop {} {\bibfield  {journal} {\bibinfo  {journal}
  {Physical review letters}\ }\textbf {\bibinfo {volume} {102}},\ \bibinfo
  {pages} {020504} (\bibinfo {year} {2009})}\BibitemShut {NoStop}%
\end{thebibliography}%

\end{document}